\documentclass[sigconf,nonacm]{acmart} 
\AtBeginDocument{%
  }

\usepackage[ruled,linesnumbered]{algorithm2e}
\usepackage{amsmath}
\usepackage{multirow}
\usepackage{amsfonts}
\usepackage{graphicx}
\usepackage{subcaption}

\usepackage{enumitem}
\setlist[itemize]{noitemsep, topsep=2pt, leftmargin=10pt}

\usepackage{xspace}
\usepackage{pifont}
\newcommand{\cmark}{\ding{51}}
\newcommand{\xmark}{\ding{55}}

\newcommand{\eg}{\textit{e.g.,}\xspace}
\newcommand{\etal}{\textit{et al.}\xspace}

\newcommand{\myparagraph}[1]{\noindent\textbf{#1}:\xspace}
\newcommand{\ts}[1]{{\scriptsize(#1)}}

\newcommand{\modelName}{\emph{PertNet}\xspace}
\newcommand{\attackName}{\emph{PosePert}\xspace}
\newcommand{\defenseName}{\emph{PoseGuard}\xspace}

\newif\ifcomments
\commentstrue

\begin{document}


\title{From Stealthy Data Fabrication to Unsafe Driving: Realistic Scenario Attacks on Collaborative Perception} 



\author{Qingzhao Zhang}
\affiliation{%
  \institution{University of Arizona}
  \city{Tucson}
  \state{Arizona}
  \country{USA}
\country{}}

\author{Runting Zhang}
\affiliation{%
  \institution{University of Michigan}
  \city{Ann Arbor}
  \state{Michigan}
  \country{USA}
\country{}}

\author{Z. Morley Mao}
\affiliation{%
  \institution{University of Michigan}
  \city{Ann Arbor}
  \state{Michigan}
  \country{USA}
\country{}}


\begin{abstract}
Collaborative perception allows connected and autonomous vehicles (CAVs) to improve perception by sharing sensory data, but it also introduces security risks from manipulated inputs. Prior work shows that attackers can spoof or remove objects by fabricating shared data, yet the practicality of such attacks in real-world driving remains unclear. Existing attacks are often detectable or evaluated in manually constructed scenarios, leaving open whether they can induce safety-critical outcomes in dynamic environments. To bridge this gap, we present a stealthy, scenario-realistic data fabrication attack that induces unsafe driving behaviors through end-to-end system effects. Instead of creating large, easily detectable anomalies, our attack subtly manipulates the poses of existing objects in shared perception results, keeping perturbations below detection thresholds. These small errors are then propagated through downstream modules, including object tracking and trajectory prediction, leading to significant deviations in predicted behaviors and ultimately unsafe driving decisions. We further design an online, scenario-aware attack framework that adapts to dynamic traffic conditions and optimizes attack strategies at runtime. Experiments on OPV2V and V2X-Real demonstrate that the attack achieves over 90\% success in inducing detection errors and triggers safety-critical behaviors, such as unnecessary hard braking, in up to 50\% of scenarios, while largely evading state-of-the-art defenses. We also propose a mitigation that focuses on detecting anomalies in localized, safety-critical regions, achieving an 80\% detection rate on the small pose perturbation compared to 11\% for the best existing methods.
\end{abstract}

%
%




\maketitle

\section{Introduction}

Connected and autonomous vehicles (CAVs) face inherent perception limitations: occlusion, limited range, and adverse conditions restrict reliable scene understanding~\cite{Pravallikaetal2024,Yazganetal2025}. \emph{Collaborative perception} addresses this by having vehicles share sensing information---point clouds, feature maps, or detections---and fuse them for broader coverage and improved accuracy~\cite{Carrilloetal2025}. This paradigm has drawn attention in both academia~\cite{thornton2025real,wan2025systematic} and industry~\cite{SAE_J3224,SUNRISE_2026,Qualcomm_Report_2023,ISTREET_2024} as a key enabler of next-generation autonomous driving.

However, sharing data introduces a security vulnerability: collaborative perception inherently trusts external inputs, creating an attack surface for \emph{data fabrication attacks}~\cite{zhang2023data,tu2021adversarial,wang2025threat,lin2025pretend} where a malicious vehicle injects crafted data to manipulate the victim’s perception. Prior work has demonstrated ghost-object creation and suppression attacks, along with defenses based on cross-vehicle consistency checking~\cite{zhang2023data,wang2025threat,zhao2024made,li2023among}.

\myparagraph{Research Gaps}
Despite these efforts, two gaps limit current understanding.
First, existing defenses~\cite{wang2025threat,zhao2024made,li2023among,hu2025cp} target large anomalies. Bounding-box methods~\cite{zhang2023data,li2023among,hallyburton2025security} threshold on significant discrepancies (insertion or removal), while feature-level defenses~\cite{wang2025threat,hu2025cp,zhao2024made} respond only when perturbations affect a large portion of the feature space. A $\sim$0.5\,m object shift introduces only minor spatial inconsistency and negligible global feature deviation, remaining below both thresholds.
Second, existing attacks~\cite{zhang2023data,tu2021adversarial,wang2025threat} are evaluated at a single frame, ignoring downstream impact. Autonomous driving is a temporal, closed-loop process where perception feeds tracking, prediction, and planning; minor errors can accumulate into unsafe actions~\cite{ma2024controlloc,muller2022physical,zhang2022adversarial,lou2024first}. Without scenario-level analysis, the true safety impact is unclear.

\myparagraph{Our Approach}
We propose a new perspective on data fabrication that jointly emphasizes \emph{stealthiness} and \emph{end-to-end impact}. Our key insight is that \emph{object pose perturbation}---subtly shifting the perceived location of an existing vehicle---is an effective attack primitive. Such perturbations evade existing defenses, yet when applied consistently over time, propagate through tracking and prediction to distort the victim’s scene understanding and cause unsafe behaviors.

\begin{figure}[t]
    \centering
    \includegraphics[width=0.47\textwidth]{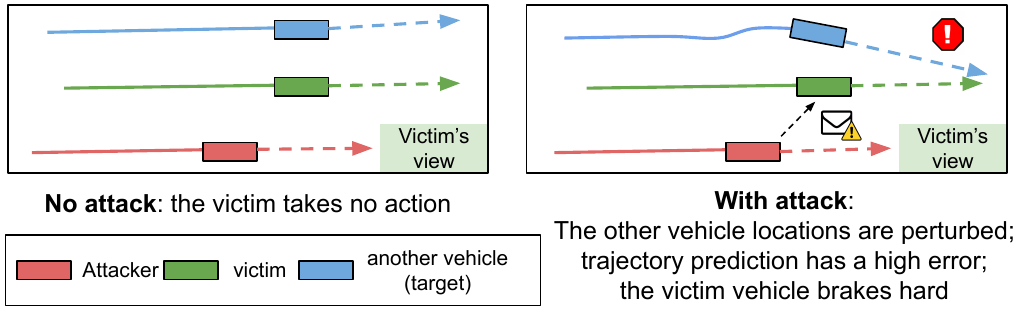}
    \caption{Illustration of the \emph{perturb-to-move-in} attack, where small per-frame shifts accumulate to induce unsafe behavior.}
    \label{fig:attack_strategy_example}
\end{figure}

To illustrate, consider the \emph{perturb-to-move-in} scenario (Figure~\ref{fig:attack_strategy_example}). The attacker injects corrupted data causing small per-frame perturbations that gradually shift a nearby vehicle toward the ego lane. Each perturbation is negligible in isolation, but their accumulation leads the victim to perceive a lane intrusion, resulting in unnecessary braking. Conversely, a \emph{perturb-to-move-out} attack shifts a genuinely approaching vehicle away, suppressing the victim’s awareness and increasing collision risk.

Realizing this attack requires \emph{precise} control of detection outputs and \emph{real-time} efficiency. We focus on \emph{intermediate fusion}~\cite{wang2020v2vnet}, where the attacker craft feature-space perturbations to induce a specific pose change, while remaining robust to benign feature interference.
We design a two-stage attack. First, \emph{scaled multi-view ray casting} provides a physics-informed initialization by synthesizing point clouds from virtual viewpoints around the target. Second, a lightweight neural network predicts per-voxel feature corrections from local context in real time, without iterative optimization or access to other vehicles’ features.
Beyond single frames, inducing unsafe behavior requires reasoning over time under partial observability. We develop a \emph{scenario-aware attack framework} following an observe--predict--plan--execute paradigm, continuously updating scene beliefs and optimizing perturbations over a temporal horizon.
Finally, we propose an \emph{object-level defense} that performs localized anomaly detection on safety-critical objects---those influencing planning decisions---improving sensitivity to small perturbations over global feature monitoring.

\myparagraph{Evaluation} We evaluate on OPV2V/OPV2V-H (simulation) and V2X-Real (real world) across four intermediate-fusion models: AttFusion~\cite{xu2022opv2v}, V2VNet~\cite{wang2020v2vnet}, CoBEVT~\cite{xu2022cobevt}, and HEAL~\cite{heal} (heterogeneous). Single-frame attacks achieve 75--94\% success (IoU$>$0.5) in $<$50\,ms, about 2$\times$ faster than PGD. Scenario attacks under a 0.5\,m per-frame bound induce up to 10\,m average trajectory deviation and reduce the predicted target-to-victim distance below 1\,m, with 25--50\% of cases entering the danger zone. Closed-loop CARLA simulation (following V2Xverse~\cite{v2xverse}) confirms that move-in attacks induce unwarranted braking in 75\% and move-out attacks increase collision risk in 40\% of scenarios. Existing defenses~\cite{zhang2023data,wang2025threat,zhao2024made,li2023among,hallyburton2025security} detect fewer than 13\% of attacks; our defense achieves 58--84\% detection at 5\% FPR by localizing anomaly detection to safety-critical objects, revealing an effectiveness--stealthiness tradeoff.

In summary, this work makes the following contributions:
\begin{itemize}
    \item We identify key limitations in existing collaborative perception security evaluations, particularly their inability to capture stealthy perturbations and their end-to-end safety impact.
    \item We propose \attackName, an \emph{object pose perturbation attack} that applies small, stealthy perturbations to manipulate perception outputs while evading existing defenses.
    \item We design a \emph{scenario-aware attack framework} that models temporal error propagation and induces unsafe behaviors by deploying \attackName over time.
    \item We develop \defenseName, an \emph{object-level anomaly detection defense} that improves detection of \attackName through localized, safety-critical analysis.
\end{itemize}
\section{Background and Related Work}
\label{sec:background}

\myparagraph{Collaborative Perception}
Collaborative perception lets CAVs overcome single-vehicle sensor limitations---occlusion, limited range, adverse weather---by sharing sensing data~\cite{lang2019pointpillars, shi2019pointrcnn, ku2018joint, alaba2022survey, li2020lidar}.
3GPP's C-V2X standard~\cite{3gpp} and investment from Huawei, Intel, Bosch, Infineon, and Qualcomm~\cite{huawei,intel,bosch,infineon,qualcomm} are driving real deployment.

Systems are classified by the level of processing before sharing:
\textit{Early fusion}~\cite{kumar2012carspeak, chen2019cooper, zhang2021emp, qiu2021autocast, chen2022cooperative, zhang2023robust} shares raw sensor data (\eg LiDAR point clouds), maximizing information at high bandwidth.
\textit{Intermediate fusion}~\cite{chen2019f, cui2022coopernaut, yuan2022keypoints, wang2020v2vnet, xu2022v2x, xu2022cobevt} shares learned feature maps, balancing bandwidth and accuracy.
\textit{Late fusion}~\cite{liu2019fusioneye, shi2022vips, song2022efficient} shares lightweight outputs such as bounding boxes, efficient but losing finer-grained information.
Intermediate fusion is the dominant paradigm and our primary focus.
While many methods assume homogeneous models across all vehicles, \emph{heterogeneous collaborative perception} allows vehicles with different model backbone or sensor modalities to collaborate via a shared feature space~\cite{heal,shao2026negocollab,zhou2026pragmatic}. We consider intermediate fusion, including this heterogeneous setting.

\begin{table}[t]
  \scriptsize
  \caption{Comparison with existing collaborative perception attacks and defenses.}
  \label{tab:related_work_comparison}
  \centering

  \begin{subtable}{\linewidth}
    \centering
    \setlength{\tabcolsep}{1pt}
    \begin{tabular}{| l | c | c | c | c | c | c |}
      \hline
      \textbf{Attacks} & \textbf{Fusion} & \textbf{Type} & \textbf{Targeted} & \textbf{Stealthy} & \textbf{Scenario} & \textbf{Computation} \\
      \hline
      Tu \etal~\cite{tu2021adversarial} & Intermediate & Untargeted & \xmark & \xmark & \xmark & Multi PGD \\
      Zhang \etal~\cite{zhang2023data} & Early/Inter. & Spoof/Remove & \cmark & \xmark & \xmark & RT + One PGD \\
      SOMBRA~\cite{wang2025threat} & Intermediate & Remove & \cmark & \xmark & \xmark & One PGD \\
      PB~\cite{lin2025pretend} & Intermediate & Spoof & \xmark & \cmark & \xmark & Multi NN \\
      \textbf{Ours} & Intermediate & Pose Perturb & \cmark & \cmark & \cmark & RT + Small NN \\
      \hline
    \end{tabular}
  \end{subtable}

  \vspace{2pt}

  \begin{subtable}{\linewidth}
    \centering
    \setlength{\tabcolsep}{1pt}
    \begin{tabular}{| l | c | c | c | c | c |}
      \hline
      \textbf{Defenses} & \textbf{Data-Level} & \textbf{Type} & \textbf{Targeted Attacks} & \textbf{Localized} & \textbf{Impact} \\
      \hline
      CAD~\cite{zhang2023data} & Box & Occupancy consistency & Spoof/Remove & \cmark & \xmark \\
      ROBOSAC~\cite{li2023among} & Box & Box-level Consensus & Untargeted Errors & \xmark & \xmark \\
      MATE~\cite{hallyburton2025security} & Box & Box Consistency & Spoof/Remove & \cmark & \xmark \\
      LUCIA~\cite{wang2025threat} & Feature Map & Feature Distances & Remove & \xmark & \xmark \\
      MADE~\cite{zhao2024made} & Feature Map & Object/Feature Distance & Untargeted Errors & \xmark & \xmark \\
      CPGuard+~\cite{hu2025cp} & Feature Map & Feature Distance & Untargeted Errors & \xmark & \xmark \\
      \textbf{Ours} & Feature Map & Object-level detection & Pose Perturb & \cmark & \cmark \\
      \hline
    \end{tabular}
  \end{subtable}

\end{table}

\myparagraph{Attacks on Collaborative Perception}
Table~\ref{tab:related_work_comparison} summarizes prior attacks and their differences from our work.

\textit{Physical-layer and single-vehicle attacks} use GPS spoofing~\cite{li2021fooling, shen2020drift}, LiDAR spoofing~\cite{jin2022pla, hallyburton2022security, li2021fooling, cao2021invisible}, or adversarial objects~\cite{tu2020physically, zhu2021can, zhang2022adversarial}, but do not exploit the collaborative channel. Our attack targets collaborative perception specifically: unlike physically influencing the victim~\cite{zhang2022adversarial,lou2024first}, it carries no physical risk; unlike manipulating the attacker's own presence~\cite{song2023discovering}, it does not require proximity.
For collaborative perception, \textit{early-fusion attacks} craft malicious point clouds to spoof or remove objects~\cite{zhang2023data}, while \textit{late-fusion attacks} trivially inject or modify shared bounding boxes~\cite{zhang2023data}.

\textit{Intermediate-fusion attacks} are the most challenging: the attacker must craft adversarial feature maps that survive fusion and detection. Tu \etal~\cite{tu2021adversarial} introduced untargeted perturbations; Zhang \etal~\cite{zhang2023data} proposed targeted spoofing/removal via PGD; SOMBRA~\cite{wang2025threat} removes objects by exploiting attention; PB~\cite{lin2025pretend} places errors in blind spots for stealthiness.

Our attack differs in three aspects: (1)~we introduce \emph{object pose perturbation}, which subtly shifts existing detections to create highly overlapping, hard-to-detect changes; (2)~we replace iterative PGD with a lightweight learned perturbation network for real-time operation; and (3)~we evaluate end-to-end through the driving pipeline, showing that small per-frame perturbations accumulate into safety-critical failures.

\myparagraph{Defenses on Collaborative Perception}
Defenses are categorized by detection level (Table~\ref{tab:related_work_comparison}).

\textit{Bounding box-level defenses} compare detections across vehicles: CAD~\cite{zhang2023data} flags occupancy-map disagreements; ROBOSAC~\cite{li2023among} samples collaborator subsets until consensus; MATE~\cite{hallyburton2025security} integrates box consistency with Bayesian trust.

\textit{Feature-level defenses} operate on shared feature maps: LUCIA~\cite{wang2025threat} computes per-agent trust from global L1 distances and reweights attention; MADE~\cite{zhao2024made} combines object- and feature-level residual analysis; CPGuard+~\cite{hu2025cp} applies feature distance metrics similarly.

Our defense performs \emph{localized} anomaly detection on \emph{safety-critical} objects, improving sensitivity to small pose perturbations that global methods miss.

\section{Motivation}
\label{sec:motivation}

Existing attacks on collaborative perception have limitations in both stealthiness and end-to-end scenario impact.

\begin{figure}[t]
    \centering
    \includegraphics[width=0.45\textwidth]{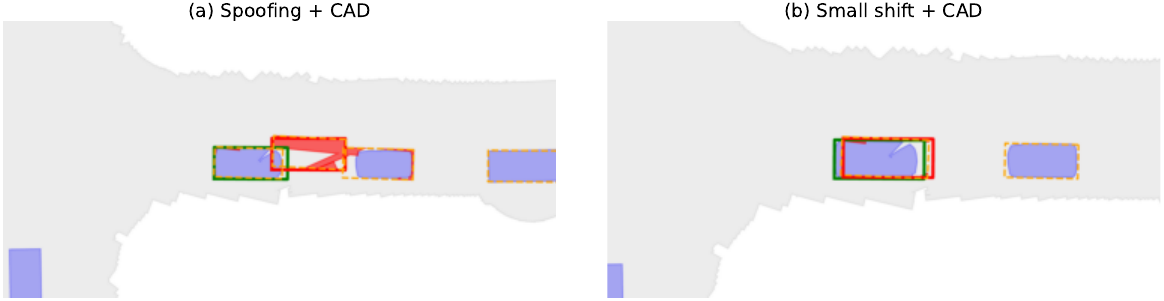}\\[1pt]
    \includegraphics[width=0.45\textwidth]{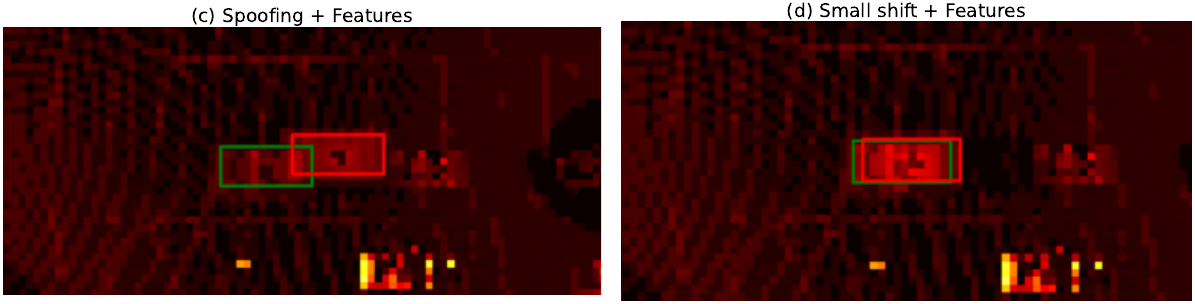}
    \caption{Examples showing that defending against small perturbations is harder (AttFusion \& OPV2V~\cite{xu2022opv2v}, Ray-Casting attack \& CAD defense~\cite{zhang2023data}).
    }
    \label{fig:stealth_demo}
\end{figure}


\myparagraph{Small perturbations evade defenses}
Defenses flag anomalies above a threshold, inherently assuming large deviations. Both bounding box-level and feature-level defenses therefore have blind spots to small, localized perturbations. We illustrate with CAD~\cite{zhang2023data} and LUCIA~\cite{wang2025threat} (Figure~\ref{fig:stealth_demo}).

\emph{Bounding box-level consistency.}
CAD~\cite{zhang2023data} constructs per-vehicle occupancy maps and flags ``conflicting regions'' where fused perception disagrees with local sensing (threshold $\sim$2.7\,m$^2$). As shown in Figure~\ref{fig:stealth_demo}(a), spoofing an object at an empty location exceeds this threshold. Shifting an existing object by 1\,m does not: the shifted and original boxes still overlap substantially (IoU$\approx$0.45 for a 4\,m car), limiting the conflict to a narrow area below the threshold. Other box-level defenses (ROBOSAC~\cite{li2023among}, MATE~\cite{hallyburton2025security}) match overlapping boxes, so small shifts are classified as normal.

\emph{Feature-level defense.}
LUCIA~\cite{wang2025threat} computes per-agent trust from global L1 feature distances and reweights attention to suppress untrusted agents. However, the feature map contains $\sim$140{,}000 spatial locations while a pose perturbation modifies only $\sim$50 voxels ($<$0.04\%). As shown in Figure~\ref{fig:stealth_demo}(d), the trust score change is negligible ($<$0.02 on average), so attention reweighting has no meaningful effect. Other feature-distance defenses share this dilution limitation (e.g., MADE~\cite{zhao2024made}).


\begin{figure}[t]
    \centering
    \includegraphics[width=0.4\textwidth]{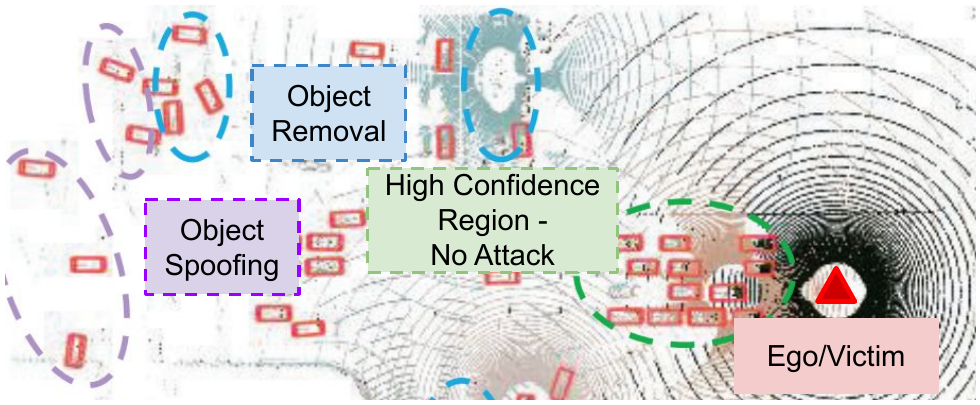}
    \caption{A blind-spot stealthy attack~\cite{lin2025pretend} (image adapted). Perception errors occur $>20$\,m from the victim.}
    \label{fig:blindspot_attack_demo}
\end{figure}

\myparagraph{Blind-spot attacks are stealthy but low-impact}
Spoofing and removal have the greatest safety impact on nearby objects, but those are directly observable by the victim, making cross-validation straightforward. Prior work (e.g., PB~\cite{lin2025pretend}) therefore places attacks in blind spots, improving stealthiness but limiting impact since affected objects are typically distant.

As illustrated in Figure~\ref{fig:blindspot_attack_demo}, the closest injected error in a representative PB scenario is $>$20\,m from the victim. Blind-spot regions are generally caused by nearby-object occlusions, so the affected targets tend to be spatially distant (e.g., multiple lanes away). Such perturbations rarely lead to safety-critical behaviors, revealing a fundamental stealthiness--impact tradeoff.


\myparagraph{Small pose perturbations can induce safety-critical errors}
Our attack resolves this tradeoff by applying small perturbations to the victim's nearby, visible objects. A single-frame error is insufficient to cause unsafe behavior, but when applied \emph{consistently across frames}, it accumulates: the tracker converges to the biased trajectory, and the predictor extrapolates it into a falsely predicted lane change (Figure~\ref{fig:attack_strategy_example}).

Related work validates this amplification mechanism: tracking hijacking~\cite{ma2024controlloc,muller2022physical} induces large association errors, and prediction attacks~\cite{zhang2022adversarial,lou2024first} show that small perturbations can significantly distort predicted motion. In collaborative perception, however, this mechanism is underexplored. Exploiting it requires \emph{online decisions}---selecting which target to perturb, determining per-frame perturbations, and adapting to evolving traffic---under substantial uncertainty, as the attacker observes the scene through its own sensors and must guess the victim's responses. We are the first to address these challenges systematically.

\section{Threat Model}
\label{sec:threat_model}

\myparagraph{System model}
We consider a collaborative perception system in which multiple vehicles share sensing data over a vehicle-to-vehicle (V2V) channel. We focus on \emph{intermediate fusion}~\cite{chen2019f, cui2022coopernaut, yuan2022keypoints, wang2020v2vnet, xu2022v2x, xu2022cobevt} as introduced in \S\ref{sec:background}. Late-fusion attacks are trivial and early fusion requires impractically high bandwidth; intermediate fusion therefore represents the most realistic and challenging setting. Further details are in \S\ref{sec:problem_formulation}. Participants may use different backbone architectures (heterogeneous setting), with adapters aligning features to a shared representation space. We assume a \emph{modular AV stack}: the perception layer is collaborative, while downstream modules (tracking, prediction, and planning) operate locally on each vehicle’s fused perception output. End-to-end driving stacks that use a single unified model are outside our current scope.

\myparagraph{Attacker capabilities}
The attacker controls at least one vehicle, with full access to its sensors, onboard software, perception stack, and communication interface. It can manipulate local sensor data, alter algorithm execution, and transmit arbitrary fabricated messages. Since the attacker participates in the collaboration with a valid identity, data fabrication is entirely local and agnostic to any authentication mechanism on the communication channel. The attacker only modifies its own data and does not intercept or tamper with communications from other vehicles.

\myparagraph{Attacker knowledge}
Following prior studies~\cite{zhang2023data,wang2025threat,zhang2023robust}, we assume white-box access to the perception model but not to downstream modules. This is natural for homogeneous intermediate fusion, where vehicles typically use the same model, so the attacker can inspect its own deployed copy. In the heterogeneous setting where vehicles use different backbones, the attacker optimizes on its local model and relies on transferability (\S\ref{sec:eval_perception_transfer}). The attacker cannot access other vehicles’ current-frame data due to sensing asynchrony and communication delays, and must rely on local observations and previously received messages. It also lacks knowledge of downstream tracking, prediction, and planning modules, and must operate under partial observability.

\myparagraph{Temporal and computational constraints}
Collaborative perception operates at 10--20\,Hz (50--100\,ms per frame). The attacker must generate and transmit fabricated data within a single sensing cycle, before fusion occurs, without access to same-frame data from other vehicles~\cite{zhang2023data,zhang2023robust}. This imposes a strict per-frame computation budget ($<$50\,ms). Our attack meets this constraint (\S\ref{sec:eval_overhead}).

\myparagraph{Attack goal}
The attacker aims to perform a \emph{targeted} attack: perturbing the perceived pose of a selected object over a temporal sequence to induce unsafe behavior in the victim vehicle. This includes unnecessary braking from a phantom lane intrusion (\emph{move-in}) or increased collision risk from suppressed awareness of a genuinely approaching vehicle (\emph{move-out}). We mainly evaluate move-in attacks because they are common in normal driving datasets, while move-out attacks require safety-critical target vehicles and are evaluated in simulation (see \S\ref{sec:eval_setup}).
\section{Design of Attack and Mitigation}
\label{sec:design}

This section presents three contributions: (1)~\attackName, a real-time pose perturbation attack on intermediate-fusion collaborative perception (\S\ref{sec:design_adv_attack}); (2)~a scenario-aware strategy that deploys \attackName over time to induce safety-critical outcomes (\S\ref{sec:attack_strategy}); and (3)~\defenseName, an object-level anomaly detection defense (\S\ref{sec:defense}). Figure~\ref{fig:overview} shows an overview.

\textbf{\attackName: Perception-Level Attack} (\S\ref{sec:design_adv_attack}).
The attacker injects crafted feature maps to perturb the perceived pose (position and orientation) of a target vehicle. Unlike prior spoofing or removal attacks~\cite{tu2021adversarial,zhang2023data}, \attackName requires precise control of detection outputs, achieved via physics-informed ray-cast initialization and a learned perturbation network (\modelName).

\textbf{Scenario-Aware Strategy} (\S\ref{sec:attack_strategy}).
To induce safety-critical effects, the attacker applies \attackName across consecutive frames, continuously updating perturbations based on new observations while accounting for scene uncertainty.

\textbf{\defenseName: Mitigation} (\S\ref{sec:defense}).
Existing defenses rely on global feature analysis and miss localized perturbations. \defenseName focuses on object-level regions to improve detection sensitivity, though the attack--defense tradeoff remains open.

\subsection{\attackName: Pose Perturbation Attack}
\label{sec:design_adv_attack}

Prior work~\cite{zhang2023data} proposed attacks to spoof or remove objects, but accurately perturbing an object's pose by a small margin (e.g., $<$1\,m) is more challenging, requiring precise control of the perturbation. We propose \attackName to bridge this gap.

\subsubsection{Problem Formulation}
\label{sec:problem_formulation}

We denote the shared data at frame $i$ from the attacker, the victim, and other benign vehicles by $f(A_i)$, $f(V_i)$, and $f(X_i^{(j)})$, $j \in \{0, 1, \dots N\}$, respectively. The format of the shared data $f(\cdot)$ depends on the fusion scheme: raw point clouds in early fusion, learned feature maps in intermediate fusion, or detection bounding boxes in late fusion. The victim fuses all shared data to produce perception results:
\begin{equation}
y_i = g\big(f(V_i),\ f(A_i),\ f(X_i^{(0)}),\ \dots,\ f(X_i^{(N)})\big).
\end{equation}
The attacker replaces its shared data $f(A_i)$ with a crafted version $f(A_i) + \delta_i$, changing the perception result to:
\begin{equation}
y_i' = g\big(f(V_i),\ f(A_i) + \delta_i,\ f(X_i^{(0)}),\ \dots,\ f(X_i^{(N)})\big).
\end{equation}
The attacker's goal is to find $\delta_i$ such that a target object's detected bounding box in $y_i'$ is shifted to a desired pose $z_t$ (\textbf{translated} and/or \textbf{rotated} from its true position), while keeping the perturbation small enough to evade defenses.

\begin{figure}[t]
    \centering
    \includegraphics[width=0.4\textwidth]{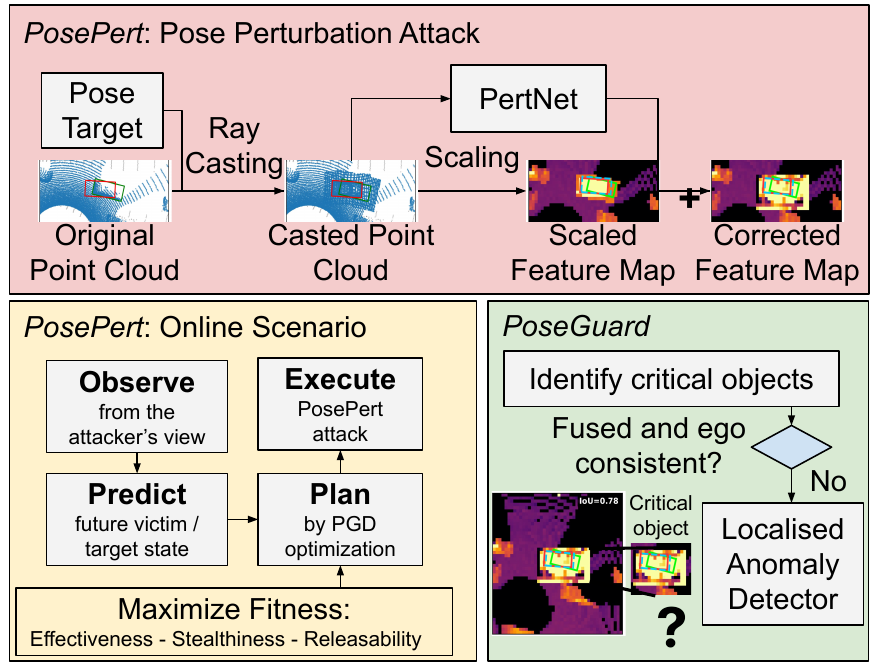}
    \caption{Overview of the proposed attack and defense.}
    \label{fig:overview}
\end{figure}

The difficulty of attacks varies across fusion schemes. In \textbf{late fusion}, attacks are trivial: the attacker injects a shifted bounding box that overrides the correct detection during NMS. In \textbf{early fusion}, the attacker must craft physically plausible point clouds at the target location. In \textbf{intermediate fusion}, the attacker must generate feature-level perturbations that induce the desired detection shift after passing through the backbone, fusion, and detection modules, while competing with benign features, making it the most challenging setting.
We focus on intermediate fusion. Figure~\ref{fig:overview} illustrates the attack pipeline.

\subsubsection{Challenges and Insights}
\label{sec:intermediate_challenges}

Intermediate-fusion pose perturbation presents unique challenges distinct from prior attacks.

\textbf{Pose perturbation $\neq$ removal $+$ spoofing.}
A natural approach is to decompose the shift into removing the object from its original location and spoofing it at the target location. However, when the shift distance is small (e.g., 1\,m for a 4\,m car), the original and target bounding boxes overlap significantly (IoU $\approx 0.4$), causing the removal and spoofing operations to conflict in the overlapping region (e.g., gradient conflicts in optimization). This fundamental conflict makes na\"ive decomposition ineffective.

\textbf{Feature scaling enhances but can break detections.}
Simply replacing the attacker's features in the target region with ray-cast features rarely shifts the detection: benign vehicles observing the object at its \emph{original} location contribute features of comparable magnitude, outvoting the attacker's shifted features.

\begin{figure}[t]
    \centering
    \includegraphics[width=0.46\textwidth]{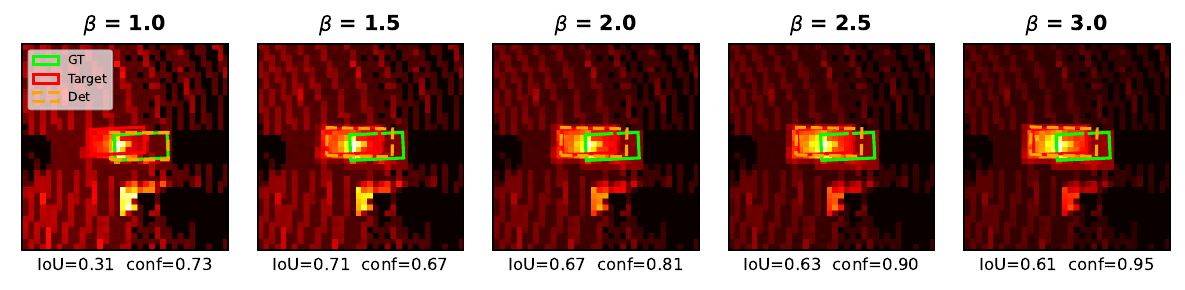}
    \caption{Effect of $\beta$ on attack success (AttFusion/OPV2V~\cite{xu2022opv2v}). IoU (0.31$\to$0.62) and confidence (0.73$\to$0.95) grow with $\beta$ scaling while excessive $\beta$ (2.5 $\to$ 3.0) could harm.}
    \label{fig:beta_scan}
\end{figure}

A natural idea is to amplify the attacker's features by a scaling factor $\beta > 1$ (multiply all feature values by $\beta$ at active locations), so that the shifted signal dominates fusion. Figure~\ref{fig:beta_scan} shows the result: at $\beta=1.0$, the detection stays near the ground truth (IoU=0.31 with target); at $\beta=1.5$, the detection shifts to the target (IoU=0.71); but at $\beta \ge 2.5$, IoU degrades (0.63$\to$0.61) even as raw confidence increases (0.90$\to$0.95).

Two properties of the perception pipeline explain this behavior. First, the pillar encoder is approximately linear in point count: the feature magnitude at each voxel scales with the number of LiDAR returns, so a ray-cast vehicle produces features of similar magnitude to a real one---insufficient to dominate over multiple benign contributions. Second, batch normalization layers in the backbone absorb moderate perturbations: their running statistics are calibrated to normal feature magnitudes, so unscaled features are normalized into the expected range regardless of spatial displacement. Scaling by $\beta \ge 1.5$ pushes features above the BN absorption threshold, allowing the shifted signal to survive through the backbone and dominate the detection head. However, excessive scaling ($\beta \ge 3$) pushes features outside the training distribution, causing the detection head to produce distorted bounding boxes. The optimal $\beta$ varies by model architecture, motivating per-model calibration.

\textbf{Perturbation generation can be learned.}
We identify geometric factors that strongly correlate with attack outcome: relative positions of original and target objects, number and positions of benign vehicles (determining the fusion ``opposition''), and local feature distribution. Crucially, these are all \emph{available to the attacker at runtime}---its own features, object locations from local perception, and other vehicles' positions from transmitted poses. Other vehicles' feature maps are \emph{not} available when transmitting the adversarial feature map. This motivates a learned perturbation generator conditioned on attacker-side information only.

\subsubsection{Feature Initialization via Multi-View Ray Casting}
\label{sec:ray_casting}

The attacker constructs a physics-informed feature initialization, which synthesizes what the target object would look like at the shifted location in three steps.

\textbf{Synthetic point cloud generation.}
We place four phantom LiDAR sources at uniform angles around the target location ($d_p = 10$\,m apart). Each uses real LiDAR sensor configurations (number of rays, angle ranges) with the origin shifted to the phantom position. Rays are cast against a 3D vehicle mesh at the target location and the estimated ground plane: intersections with the mesh produce car surface points, while misses hitting the ground produce surrounding ground points, mimicking a natural LiDAR pattern.

\textbf{Feature extraction.}
The synthetic point cloud is processed through the same feature extraction pipeline used by the perception model, yielding the spoofed feature map $f_{\text{spoof}}$. This feature map represents the attacker's contribution \emph{as if} the object were physically present at the target location.

\textbf{Scaled replacement.}
The attacker replaces its features in the target region with a scaled version of the spoofed features:
\begin{equation}
\label{eq:base_pert}
f_{\text{base}} = \beta \cdot f_{\text{spoof}}
\end{equation}
where $\beta \ge 1$ is the scaling factor calibrated per model. The base perturbation relative to the original features is $\delta_{\text{base}} = f_{\text{base}} - f(A_i)$. This initialization provides a strong directional signal for the detection shift, grounded in realistic point cloud geometry.

\subsubsection{Learned Perturbation Network}
\label{sec:pertnet}

The ray-cast initialization is physically grounded but unaware of the fusion model's learned behavior and scene-specific geometry. We therefore train a lightweight convolutional network $\mathcal{G}_\theta$ that predicts a per-voxel correction on top of the base.

\textbf{Network input.}
The network operates on a local crop of the feature map centered on the target region and receives three inputs: (1) the attacker's original features $f_{\text{orig}}$, representing the current feature distribution; (2) the ray-cast feature difference $f_{\text{diff}} = f_{\text{spoof}} - f_{\text{orig}}$, encoding the direction of the desired feature change; and (3) a geometric encoding $\mathbf{e}_{\text{geo}}$ with per-voxel channels capturing the offset from the original and target bbox centers, binary masks of original and target footprints, and the bearing angle and inverse distance to each benign vehicle's position. All inputs are available to the attacker without requiring other vehicles' feature maps.

\textbf{Network architecture.}
The network consists of three convolutional layers (with $3 \times 3$ kernels) followed by a $1 \times 1$ output layer that produces a correction $\delta_{\text{corr}} \in \mathbb{R}^{C \times h \times w}$ matching the feature dimensionality at each voxel. This produces spatially varying corrections---each voxel receives a different correction depending on whether it contains car, ground, or empty features---without global pooling that would lose spatial specificity.

\textbf{Combined perturbation.}
The final transmitted feature map combines the scaled base with the bounded correction:
\begin{equation}
\label{eq:combined_pert}
f_{\text{attack}} = f(A_i) + \delta_{\text{base}} + \text{clamp}(\delta_{\text{corr}},\ -\epsilon,\ \epsilon)
\end{equation}
where $\epsilon$ bounds the correction magnitude independently of the base. This ensures the learned correction refines the initialization rather than dominating or undermining it.

\textbf{Training objective.}
The network is trained end-to-end through the frozen perception model. Each training sample provides the attacker's features, spoofed features, and geometric context from the training split of attack scenarios. The perturbed feature map is passed through the victim's full perception pipeline to produce detection proposals. The loss maximizes IoU with the target box while preserving detection confidence:
\begin{equation}
\label{eq:pertnet_loss}
\mathcal{L} = \sum_{\substack{z \in \mathcal{P},\ \text{IoU}(z, z_o) > 0, \\ \text{IoU}(z, z_t) > 0,\ \sigma > \epsilon_\sigma}} \log(1 - \text{IoU}(z, z_t)) + \lambda \cdot \max(0,\ \sigma_0 - \sigma)
\end{equation}
where $z$ denotes detection proposals with confidence $\sigma$, $z_o$ and $z_t$ are the original and target bounding boxes, $\sigma_0$ is the pre-attack confidence, and $\lambda$ balances the objectives. The loss selects proposals that overlap both the original and target boxes---the proposals most likely to be ``perturbable''---and pushes them toward the target while penalizing confidence drop.

\textbf{Inference-time execution.}
At inference, the full perception attack executes in a single forward pass: ray casting ($\sim$20\,ms), feature extraction, and \modelName correction ($<$1\,ms). The total latency is well within the $\sim$100\,ms frame interval of typical LiDAR systems. See \S\ref{sec:eval_overhead} for efficiency analysis.

\subsection{Scenario-Aware Strategy for \attackName}
\label{sec:attack_strategy}

\attackName (\S\ref{sec:design_adv_attack}) operates at individual frames. To induce safety-critical outcomes, the attacker must orchestrate attacks across a temporal sequence, adapting to the evolving traffic scene.

\subsubsection{Problem Formulation}

We formulate the attack using the ``perturb to move in'' scenario (Figure~\ref{fig:attack_strategy_example}) as the primary example, where the attacker shifts the perceived location of a target vehicle driving alongside the victim to simulate a lane-changing maneuver, causing the victim to hard-brake. The dual ``perturb to move out'' scenario reverses the direction: a genuinely approaching vehicle is shifted away, suppressing the victim's awareness and increasing collision risk. Both share the same formulation below, differing only in the sign of the effectiveness objective. An attack strategy consists of:
\begin{itemize}
    \item A \textbf{victim vehicle} $V$ whose behavior the attacker aims to influence.
    \item A \textbf{target vehicle} $Q$ whose perceived location will be shifted.
    \item An \textbf{adversarial trajectory} $T_{\text{adv}} = \{(x_i, y_i)\}_{i=1}^{K}$ specifying the desired perceived position of $Q$ at each of the $K$ attack frames.
\end{itemize}

The problem is to choose the adversarial trajectory and achieve it through \attackName. It should satisfy three properties:
\begin{itemize}
    \item \textbf{Effectiveness}: For move-in, the predicted trajectory of $Q$ should come within a distance threshold $d_s$ of the victim's planned path, triggering a defensive hard brake. For move-out, the predicted trajectory should \emph{exceed} $d_s$, masking a genuine interaction and increasing collision risk.
    \item \textbf{Stealthiness}: The per-frame locational perturbation $\|T_{\text{adv}}^{(i)} - T_Q^{(i)}\|$ must remain within a bound $\epsilon_{\text{loc}}$ (we use $\epsilon_{\text{loc}} = 0.5$\,m) to stay below anomaly detection thresholds.
    \item \textbf{Realizability}: The adversarial trajectory must be achievable by the perception attack---the target position at each frame must lie within the attacker's feature map coverage.
\end{itemize}

\subsubsection{Challenges and Our Approach}
\label{sec:scenario_challenges}

Orchestrating perception attacks into safety-critical scenarios requires bridging several gaps:

\textbf{Unknown downstream models.}
The attacker does not know the victim’s tracking/prediction model. We adopt a \emph{surrogate model} approach, optimizing the adversarial trajectory against a local predictor. Our evaluation (\S\ref{sec:eval_scenario}) shows that such perturbations transfer effectively across different models, achieving comparable danger rates, as the attack primarily exploits tracking corruption rather than predictor-specific behavior.

\textbf{White-box vs.\ query-access optimization.}
When the victim’s model parameters or gradients are unavailable (or non-differentiable), we use \emph{query-based} optimization via finite-difference gradient estimation: each parameter is perturbed by $\pm\epsilon$ and evaluated, requiring $2d$ queries per iteration.

\textbf{Target vehicle selection.}
The attacker pre-analyzes the scene from its local observations to select the most effective target. Ideal targets are vehicles near the victim, typically in adjacent lanes, where small lateral perturbations can plausibly indicate a lane change. We rank candidates by proximity and lateral offset, selecting the one most likely to induce a safety-critical prediction.

\textbf{Unknown scene future.}
The attacker cannot know how the scene will evolve during the multi-frame attack window. At each frame, the attacker observes the current scene, predicts near-term evolution using the surrogate model, re-optimizes the adversarial trajectory, and executes the next attack. This online replanning adapts to dynamic conditions.

\textbf{Perception attack imprecision.}
The perception attack does not perfectly realize the desired perturbation at each frame due to benign feature interactions and pipeline variability. Rather than explicitly modeling this imprecision, we rely on the replanning loop to correct deviations: individual steps may deviate, but the overall trajectory is steered toward safety-critical outcomes.

\subsubsection{Attack Quality Fitness Function}

The attacker evaluates candidate attack strategies using a fitness function that balances effectiveness, stealthiness, and realizability:

\begin{equation}
\label{eq:fitness}
\texttt{AttackQuality}(T_{\text{adv}}) = \begin{cases}
l_e & \text{if } l_s > 0 \text{ and } l_r > 0 \\
0 & \text{otherwise}
\end{cases}
\end{equation}
where the three components are:
\begin{itemize}
    \item $l_e = -\log d(\texttt{Pred}(T_{\text{adv}}^{1:K}),\ \texttt{Pred}(T_V^{1:K}))$: the effectiveness score, for move-in scenarios, measuring how close the predicted trajectory of $Q$ (given all $K$ adversarial observations) comes to the victim's predicted path. For the move-out scenario the sign is reversed ($l_e = +\log d(\cdot)$), rewarding increased distance.
    \item $l_s$: the stealthiness score, a binary check that is positive if and only if all per-frame perturbation magnitudes remain within a threshold, e.g., $\epsilon_{\text{loc}} = 0.5$\,m.
    \item $l_r$: the realizability score, a binary check that is positive if and only if all adversarial positions fall within the attacker's feature map coverage (${\sim}$40\,m range). Otherwise the malicious data cannot reach the victim's data fusion.
\end{itemize}

\subsubsection{End-to-End Online Attack}
\label{sec:optimization}

\begin{algorithm}[t]
\small
\SetAlgoLined
\KwIn{The attacker identified a victim vehicle $V$. The attack lasts for $K$ frames and optimizes attack impact on the final frame $K$. The attacker maintains observed or predicted trajectories of other vehicles, denoted by a set $T$ where $T_i$ is the trajectory of vehicle $i$ over $K$ frames.}

\SetKwFunction{FMain}{OnlineAttack}
\SetKwProg{Fn}{Function}{:}{}
\Fn{\FMain{}}{
    $T_{\text{adv}} \gets \operatorname*{argmax}_{T_i \in T} \texttt{AttackQuality}(T_i)$\;
    \For {Frame $k = 1, \ldots, K$} {
        $\texttt{PerceptionAttack}(T_{\text{adv}}^{(k)})$\;
        $T \gets \texttt{UpdateSceneKnowledge}(T)$
        $T_{\text{adv}}^{(k:K)} \gets \texttt{PGDUpdate}(T_{\text{adv}}, k{:}K)$\;
    }
}

\SetKwFunction{FMain}{PGDUpdate}
\SetKwProg{Fn}{Function}{:}{}
\Fn{\FMain{$T_{\text{adv}}$, $i{:}j$}}{
    $\lambda \gets \max \{ \lambda\ |\ \texttt{AttackQuality}(\lambda(T_{\text{adv}} - T_Q)+T_Q) > 0 \}$\;
    $T_{\text{adv}} \gets \lambda(T_{\text{adv}} - T_Q)+T_Q$\;
    \Return $\texttt{AdamOptimizer}(\texttt{AttackQuality}(T_{\text{adv}}), T_{\text{adv}}^{i:j})$\;
}
 \caption{End-to-end online attack algorithm.}
 \label{alg:online_attack}
\end{algorithm}

As on-road traffic is highly dynamic, the attacker must continuously adapt its strategy. Algorithm~\ref{alg:online_attack} presents the end-to-end attack, structured as an \emph{observe--predict--plan--execute} loop that repeats at each frame.

\textbf{Observe.}
The attacker uses its onboard perception and tracking stack to maintain trajectory estimates $T$ for all surrounding vehicles. Before the attack begins, the attacker identifies the victim $V$ and selects the target vehicle $Q$ that maximizes the attack quality fitness among nearby candidates (e.g., the vehicle closest to the victim in an adjacent lane). At each subsequent frame $k$, the attacker updates $T$ with the latest sensing results, correcting for prediction drift and newly observed vehicle behavior.

\textbf{Predict.}
Using a surrogate trajectory prediction model, the attacker forecasts how the scene will evolve over the remaining $K - k$ frames. This prediction informs the optimization: the attacker evaluates how candidate adversarial trajectories would affect the victim's predicted response, including whether the target would appear to enter the victim's lane.

\textbf{Plan.}
The attacker optimizes the remaining adversarial trajectory $T_{\text{adv}}^{(k:K)}$ via Projected Gradient Descent (PGD). The \texttt{PGDUpdate} function first projects the trajectory to the feasible set by scaling toward the true trajectory $T_Q$ until stealthiness and realizability constraints are satisfied, then applies Adam optimization to maximize the attack quality score. Only future frames are optimized---past perception attacks are already committed and cannot be revised.

\textbf{Execute.}
The attacker launches the pose perturbation perception attack (\S\ref{sec:design_adv_attack}) to realize the current frame's adversarial position $T_{\text{adv}}^{(k)}$, transmitting the crafted feature map to the victim. The loop then advances to the next frame.

This closed-loop design handles the uncertainties identified in \S\ref{sec:scenario_challenges}: online replanning adapts to scene changes, the PGD projection enforces stealthiness constraints, and the surrogate predictor provides gradient signal when the victim's model is unknown.

\subsection{\defenseName: Localized Anomaly Detection}
\label{sec:defense}

We present a mitigation that concentrates anomaly detection on individual detected objects, significantly increasing sensitivity to localized attacks.

\subsubsection{Insight: Why Object-Level Detection is Necessary}

\begin{figure}[t]
    \centering
    \includegraphics[width=0.30\textwidth]{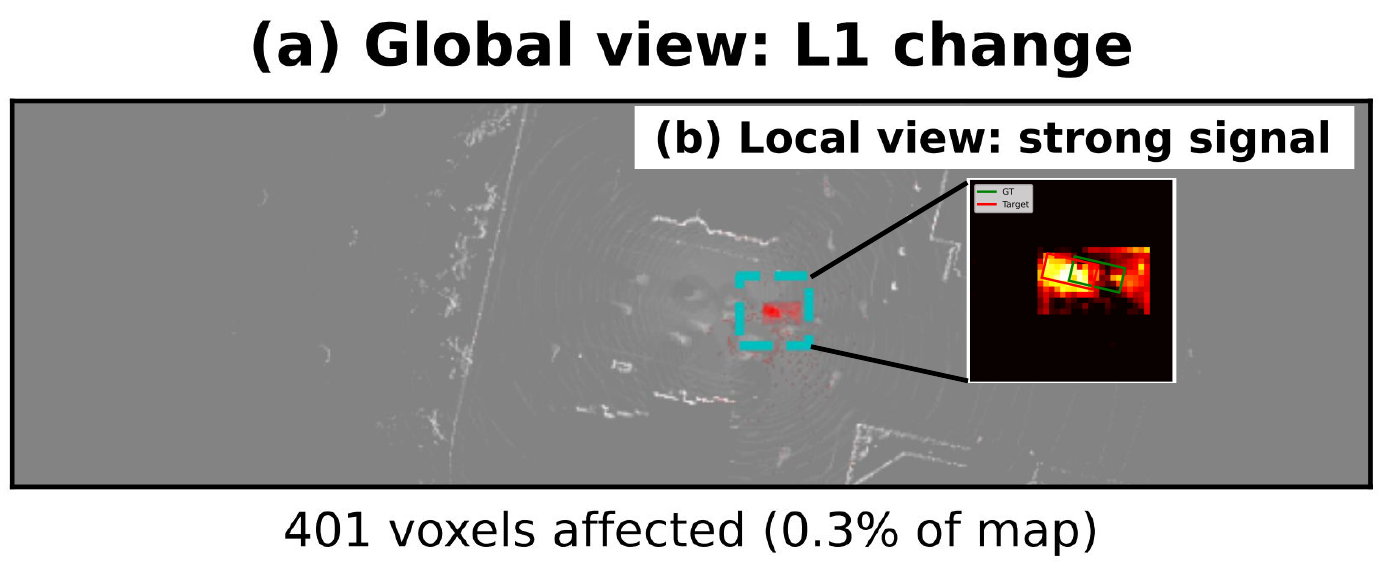}
    \caption{Signal dilution in global vs.\ local anomaly detection (AttFusion/OPV2V~\cite{xu2022opv2v}, LUCIA~\cite{wang2025threat}).}
    \label{fig:lucia_dilution}
\end{figure}

Consider a feature map of size $C \times H \times W$ with $H \times W \approx 140{,}000$ spatial locations. A pose perturbation attack modifies features in a region of approximately $h \times w \approx 400$ voxels---less than 0.3\% of the total feature map. As shown in Figure~\ref{fig:lucia_dilution}, global anomaly metrics (e.g., the L1 distance used by LUCIA~\cite{wang2025threat}) produce a signal-to-noise ratio near 1.0$\times$: the perturbation's L1 contribution is diluted across the entire map. This dilution affects not only detection but also LUCIA's attention-reweighting mechanism: because the trust score changes by only $\sim$0.015 under attack (from 0.509 to 0.494 on average), the resulting attention adjustment has negligible effect on the fusion output. In contrast, cropping the anomaly computation to the bounding box region concentrates the signal to 3--9$\times$.

This motivates a pipeline that (1) identifies \emph{which} objects to scrutinize and (2) performs \emph{per-object} anomaly detection using local feature crops. Recovery strategies are orthogonal and left to future work.

\subsubsection{Detection Pipeline}

\defenseName operates in three stages, progressively filtering the set of objects that require deeper analysis:

\textbf{Stage 1: Safety-critical object identification.}
Not all detected objects require anomaly checking. The defender first identifies objects whose predicted future trajectories come within a safety threshold $d_s$ of the ego vehicle's planned path. Only these \emph{safety-critical} objects proceed to the next stage, avoiding unnecessary computation on distant or irrelevant detections. One typical implementation is the trial-and-test approach: perturbing the bounding box by a small margin (e.g., 0.5\,m) and observing whether that could affect driving decisions (e.g., trajectory prediction on perturbed trajectories ever interact with the future path).

\textbf{Stage 2: Fused-vs-ego disagreement filter.}
For each critical object, the defender compares the \emph{fused} detection (from collaborative perception) with the \emph{ego-only} detection (from the ego's own sensors alone). If the two detections agree (IoU $> \tau_{\text{iou}}$ and center distance $< \tau_d$), the object is likely genuine and no further analysis is needed. Only objects where fused and ego detections \emph{disagree} are flagged for anomaly checking.

\textbf{Stage 3: Object-level anomaly detection.}
For each disagreeing critical object, the defender extracts the \emph{local} feature crop from each vehicle's spatial feature map in the bounding box region. It then L2-normalizes each agent's crop and computes pairwise L1 distances between all agents (following LUCIA's methodology~\cite{wang2025threat} but applied per-object rather than globally). An anomaly score exceeding a calibrated threshold $\tau_a$ (set at the 95th percentile of clean cases) flags the object as potentially compromised.

\subsubsection{Discussion}

The object-level defense creates a fundamental trade-off for the attacker: using large perturbations ($\beta \gg 1$) increases the probability of shifting the detection but also increases the anomaly score in the local region. Conversely, small perturbations ($\beta \approx 1$) evade detection but fail to shift the object reliably. See \S\ref{sec:eval_defense_resilience} for details.
\section{Evaluation}
\label{sec:evaluation}

We evaluate our proposed attack \attackName at the perception level (\S\ref{sec:eval_perception}) and the scenario level (\S\ref{sec:eval_scenario}), measuring both attack effectiveness and defense resilience across multiple models and datasets.

\subsection{Experimental Setup}
\label{sec:eval_setup}

\myparagraph{Datasets}
We evaluate on two collaborative perception datasets.
\begin{itemize}
    \item \textbf{OPV2V}~\cite{xu2022opv2v} is a large-scale simulated multi-vehicle dataset generated via co-simulation of CARLA~\cite{carla} and SUMO~\cite{sumo}, providing synchronized LiDAR point clouds and ground-truth annotations across 2--5 collaborative vehicles. We also use \textbf{OPV2V-H}~\cite{heal} (augmented from OPV2V) for the heterogeneous collaboration.
    \item \textbf{V2X-Real}~\cite{xiang2024v2x} is a real-world vehicle-to-vehicle dataset collected from instrumented vehicles with LiDAR sensors and infrastructure nodes.
\end{itemize}

\myparagraph{Test cases}
We construct two types of test cases. We select test cases with a filter that ensures data quality. Detailed filtering criteria are in Appendix~\ref{sec:appendix_details}.
\begin{itemize}
    \item \textbf{Perception test cases} (300/200 from OPV2V/V2X-Real) each specify an attacker, victim, target object, and a random shift---distance in $[0.5, 2.0]$\,m, towards a random direction, with a small random rotation in $[-10^\circ, 10^\circ)$.
    \item \textbf{Scenario test cases} (102/45 from OPV2V/V2X-Real) each specify a multi-frame driving scenario with a target vehicle approaching the victim. These cases evaluate the \emph{move-in} attack on logged data, turning normal driving in the logged data into unexpected braking scenarios. Cases are filtered to ensure the initial target-to-victim distance is within 10\,m. 102 of 120 OPV2V scenarios (85\%) remain after filtering, showing that vulnerable scenarios for move-in attacks are not sparse in existing datasets.
    \item \textbf{Closed-loop simulation} (Section~\ref{sec:eval_closed_loop}) evaluates both move-in (12 cases) and move-out attacks (10 cases) in a reactive driving stack on CARLA. In the move-out scenarios, the simulation records increased collision risk, including actual collisions.
\end{itemize}

\myparagraph{Models}
We evaluate the following intermediate-fusion architectures, the first three implemented through OpenCOOD~\cite{xu2022opv2v}:
\begin{itemize}
    \item \textbf{AttFusion}~\cite{xu2022opv2v}: Attention-based feature fusion. Tested on both OPV2V and V2X-Real.
    \item \textbf{V2VNet}~\cite{wang2020v2vnet}: Spatial-aware message passing with learned communication topology. Tested on OPV2V.
    \item \textbf{CoBEVT}~\cite{xu2022cobevt}: Transformer-based multi-agent fusion with fused axial attention. Tested on OPV2V.
    \item \textbf{HEAL}~\cite{heal}: Heterogeneous collaborative perception with learnable feature alignment across different backbone architectures. Tested on OPV2V-H for cross-modality transfer evaluation.
\end{itemize}
All models above use the PointPillars~\cite{lang2019pointpillars} backbone, a commonly used encoder for point cloud to spatial feature map conversion. For scenario evaluation, the downstream autonomous driving stack uses AB3DMOT~\cite{weng2020ab3dmot} for multi-object tracking and GRIP++~\cite{li2019grip++} for trajectory prediction (20-frame observation, 20-frame prediction at 10\,Hz). We also evaluate with Trajectron++~\cite{salzmann2020trajectron++} for transferability. These are widely-used modular AV components (e.g.,~\cite{v2xverse}).

\myparagraph{Attack configuration}
For perception attacks, multi-view ray casting uses 4 phantom LiDARs with Open3D~\cite{zhou2018open3d}. The scaling factor $\beta$ is selected per model via grid search ($\beta=2.0$ for PP-AttFusion and CoBEVT, $\beta=3.0$ for V2VNet). \modelName is trained per model on 1{,}000 ray-cast samples from the train split.
For scenario attacks, the attacker applies 5 sign-based PGD iterations per frame to optimize a perturbation across 3 frames ($K=3$) with perturbation bound $\epsilon_{\text{loc}} = 0.5$\,m. By default the trajectory predictor observes 2\,s and predicts 2\,s.

\myparagraph{Attack baseline}
We adopt the PGD-based attack from Zhang \etal~\cite{zhang2023data} as a baseline. It optimizes feature-map perturbations online using one PGD step per frame and reuses the perturbation across frames. For a fair comparison, we adapt it to use the same objective function as \modelName.

\myparagraph{Defenses}
We evaluate the following defenses against our attack:
\begin{itemize}
    \item \textbf{CAD}~\cite{zhang2023data}: Occupancy consistency check. Builds per-vehicle occupancy maps and flags detections overlapping free space.
    \item \textbf{LUCIA}~\cite{wang2025threat}: Computes per-agent trust scores from global L1 feature distances and uses them to reweight attention during fusion, serving as both detector and mitigation.
    \item \textbf{MADE}~\cite{zhao2024made}: Residual feature analysis. Measures leave-one-out residual features with autoencoder reconstruction.
    \item \textbf{ROBOSAC}~\cite{li2023among}: Consensus-based vehicle exclusion. Repeatedly samples subsets of collaborators until perception results converge, excluding outliers.
    \item \textbf{MATE}~\cite{hallyburton2025security}: Trust-based anomaly detection. Maintains per-agent trust scores updated by detection consistency over time.
\end{itemize}
Implementation details and threshold calibration are in Appendix~\ref{sec:appendix_details}.

\subsection{Perception Attack Results}
\label{sec:eval_perception}

We evaluate the pose perturbation attack on single frames, with factor analysis and ablation.

\subsubsection{Effectiveness}
\label{sec:eval_effectiveness}

\begin{figure*}[t]
\begin{minipage}[t]{0.60\textwidth}
\vspace{0pt}
  \small
  \captionof{table}{Perception attack effectiveness and defense detection rates.}
  \label{tab:evaluation_perception_results}
  \setlength{\tabcolsep}{2pt}
  \begin{tabular}{| l | l | c | c | c | c | c | c | c | c | c |}
    \noalign{\global\arrayrulewidth1pt}\hline\noalign{\global\arrayrulewidth0.4pt}
    \multirow{2}{*}{\textbf{Setting}} & \multirow{2}{*}{\textbf{Attack}} & \multicolumn{5}{c|}{\textbf{Attack Effectiveness}} & \multicolumn{4}{c|}{\textbf{Defense TPR@5\%FPR}} \\
    \cline{3-11}
    & & \%Imp & \%S(.5) & \%S(.7) & IoU ($\Delta$) & Conf ($\Delta$) & CAD & LUCIA & MADE & Ours \\
    \noalign{\global\arrayrulewidth1pt}\hline\noalign{\global\arrayrulewidth0.4pt}
    \multirow{4}{*}{\shortstack[l]{AttFusion\\OPV2V}}
    & PGD & 48.2 & 14.0 & 0.3 & .32\ts{-.00} & .62\ts{-.03} & \textbf{23.5} & 4.7 & 6.3 & 5.0 \\
    & Ray-cast & 88.0 & 41.7 & 16.0 & .42\ts{+.10} & .55\ts{-.10} & \textbf{14.0} & 4.3 & 7.0 & 9.3 \\
    & +$\beta$=2.0 & 86.7 & 59.3 & 28.7 & .50\ts{+.18} & .54\ts{-.11} & 34.7 & 5.8 & 6.7 & \textbf{82.3} \\
    & +\modelName & \textbf{94.0} & \textbf{75.0} & \textbf{34.0} & \textbf{.59}\ts{+.27} & \textbf{.56}\ts{-.09} & 50.3 & 7.7 & 8.3 & \textbf{96.0} \\
    \hline
    \multirow{4}{*}{\shortstack[l]{V2VNet\\OPV2V}}
    & PGD & 36.5 & 9.0 & 0.0 & .30\ts{-.02} & .76\ts{-.06} & \textbf{18.0} & 4.3 & 5.7 & 5.3 \\
    & Ray-cast & 98.0 & 67.7 & 25.3 & .57\ts{+.25} & .74\ts{-.07} & \textbf{22.4} & 5.6 & 7.3 & 15.0 \\
    & +$\beta$=3.0 & 97.3 & 90.3 & 50.0 & .66\ts{+.34} & .72\ts{-.09} & 48.1 & 8.4 & 8.7 & \textbf{95.0} \\
    & +\modelName & \textbf{98.0} & \textbf{94.0} & \textbf{57.3} & \textbf{.70}\ts{+.38} & \textbf{.77}\ts{-.04} & 46.7 & 10.3 & 9.3 & \textbf{99.3} \\
    \hline
    \multirow{4}{*}{\shortstack[l]{CoBEVT\\OPV2V}}
    & PGD & 38.5 & 15.1 & 0.7 & .31\ts{-.01} & .61\ts{-.06} & \textbf{20.0} & 4.3 & 6.0 & 6.0 \\
    & Ray-cast & 88.3 & 62.0 & 38.0 & .53\ts{+.21} & .53\ts{-.14} & \textbf{18.6} & 4.7 & 7.7 & 10.0 \\
    & +$\beta$=2.0 & 92.3 & 77.3 & 54.7 & .61\ts{+.29} & .52\ts{-.15} & 35.2 & 5.2 & 8.0 & \textbf{85.0} \\
    & +\modelName & \textbf{96.3} & \textbf{93.3} & \textbf{54.7} & \textbf{.69}\ts{+.37} & \textbf{.56}\ts{-.11} & 72.7 & 6.7 & 9.0 & \textbf{98.0} \\
    \hline
    \multirow{4}{*}{\shortstack[l]{AttFusion\\V2X-Real}}
    & PGD & 58.0 & 18.5 & 3.0 & .31\ts{+.04} & .31\ts{-.06} & \textbf{9.0} & 5.0 & 4.5 & 8.5 \\
    & Ray-cast & 68.5 & 58.0 & 47.0 & .48\ts{+.23} & .40\ts{-.09} & \textbf{12.0} & 5.5 & 5.5 & 9.0 \\
    & +$\beta$=1.2 & 72.0 & 59.0 & 48.0 & .50\ts{+.22} & .40\ts{-.08} & 15.0 & 6.0 & 6.0 & \textbf{68.0} \\
    & +\modelName & \textbf{85.5} & \textbf{76.0} & \textbf{46.0} & \textbf{.59}\ts{+.33} & \textbf{.45}\ts{-.04} & 21.5 & 6.5 & 7.0 & \textbf{80.5} \\
    \noalign{\global\arrayrulewidth1pt}\hline\noalign{\global\arrayrulewidth0.4pt}
  \end{tabular}

\end{minipage}%
\hfill
\begin{minipage}[t]{0.40\textwidth}
\vspace{0pt}
\centering
\includegraphics[width=\textwidth]{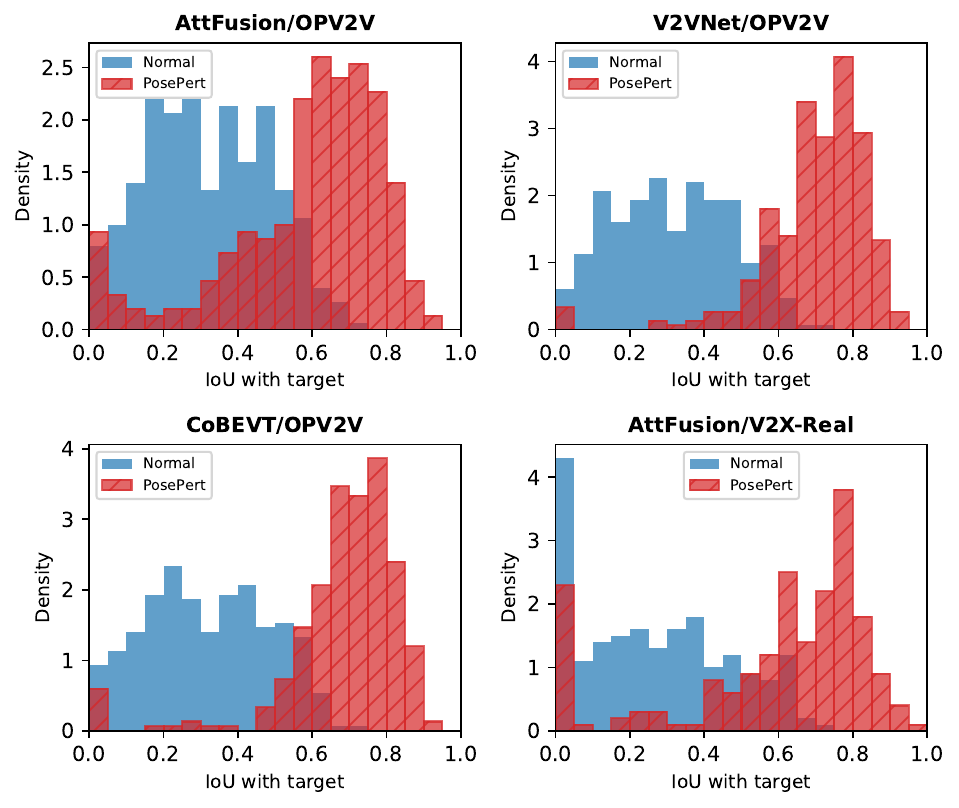}
\captionof{figure}{IoU distributions before (blue) and after (red) the full attack.}
\label{fig:iou_distributions}
\end{minipage}
\end{figure*}

Attack effectiveness metrics (Table~\ref{tab:evaluation_perception_results}):
\begin{itemize}
    \item \textbf{\%Improved}: Fraction of cases where IoU(detection, target) increases under attack.
    \item \textbf{\%Success(IoU$>$0.5)}: Fraction of cases achieving IoU $>$ 0.5 with the target bounding box. Indicates accurate shifting.
    \item \textbf{\%Success(IoU$>$0.7)}: Fraction of cases achieving IoU $>$ 0.7 with the target. Indicates precise shifting.
    \item \textbf{Avg(IoU)}: Average IoU between the detection \& the target box.
    \item \textbf{Avg(Confidence)}: Average detection confidence after attack.
\end{itemize}

\begin{figure*}[t]
\begin{minipage}[t]{0.34\textwidth}
\vspace{0pt}
\centering
\includegraphics[width=\textwidth]{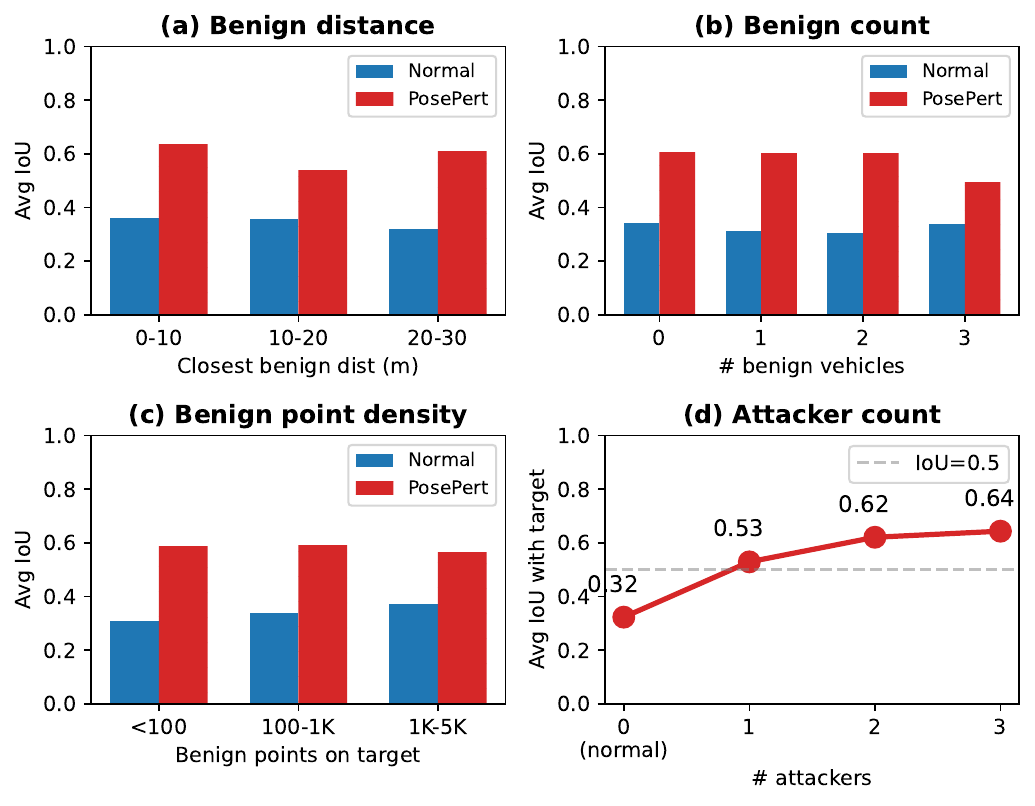}
\captionof{figure}{Factors affecting attack success.}
\label{fig:factor_distance}
\end{minipage}%
\hfill
\begin{minipage}[t]{0.30\textwidth}
\vspace{0pt}
\centering
\includegraphics[width=\textwidth]{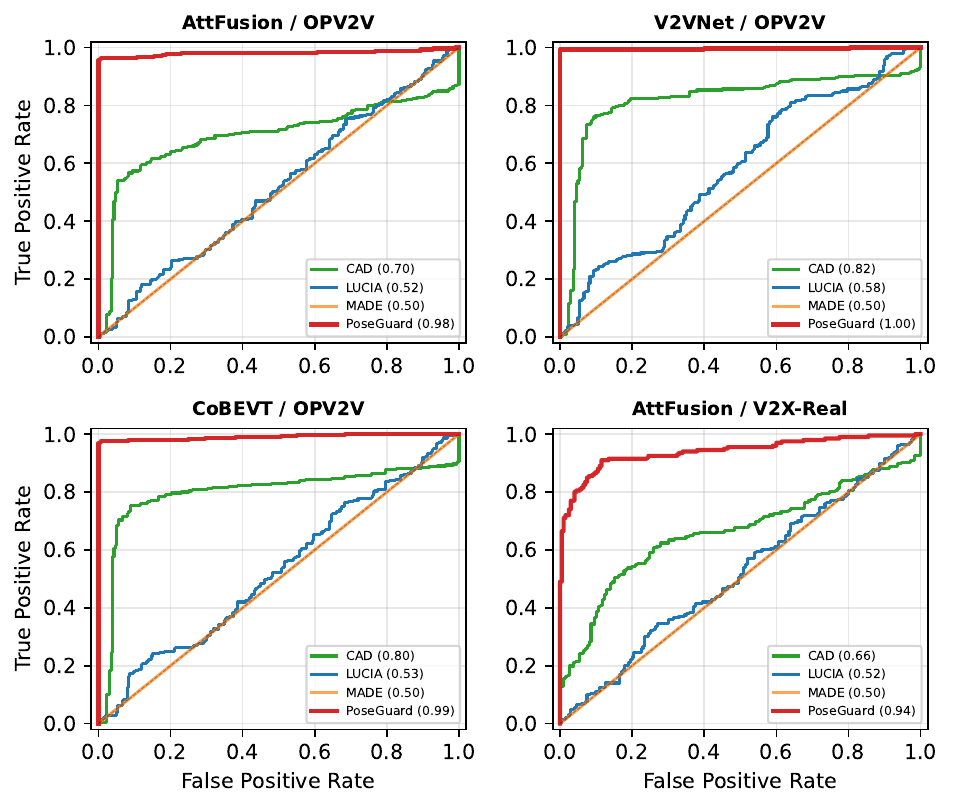}
\captionof{figure}{Defense ROC curves. \defenseName AUC$>$0.98; CAD 0.70--0.82.}
\label{fig:defense_roc}
\end{minipage}%
\hfill
\begin{minipage}[t]{0.34\textwidth}
\vspace{0pt}
\centering
\includegraphics[width=\textwidth]{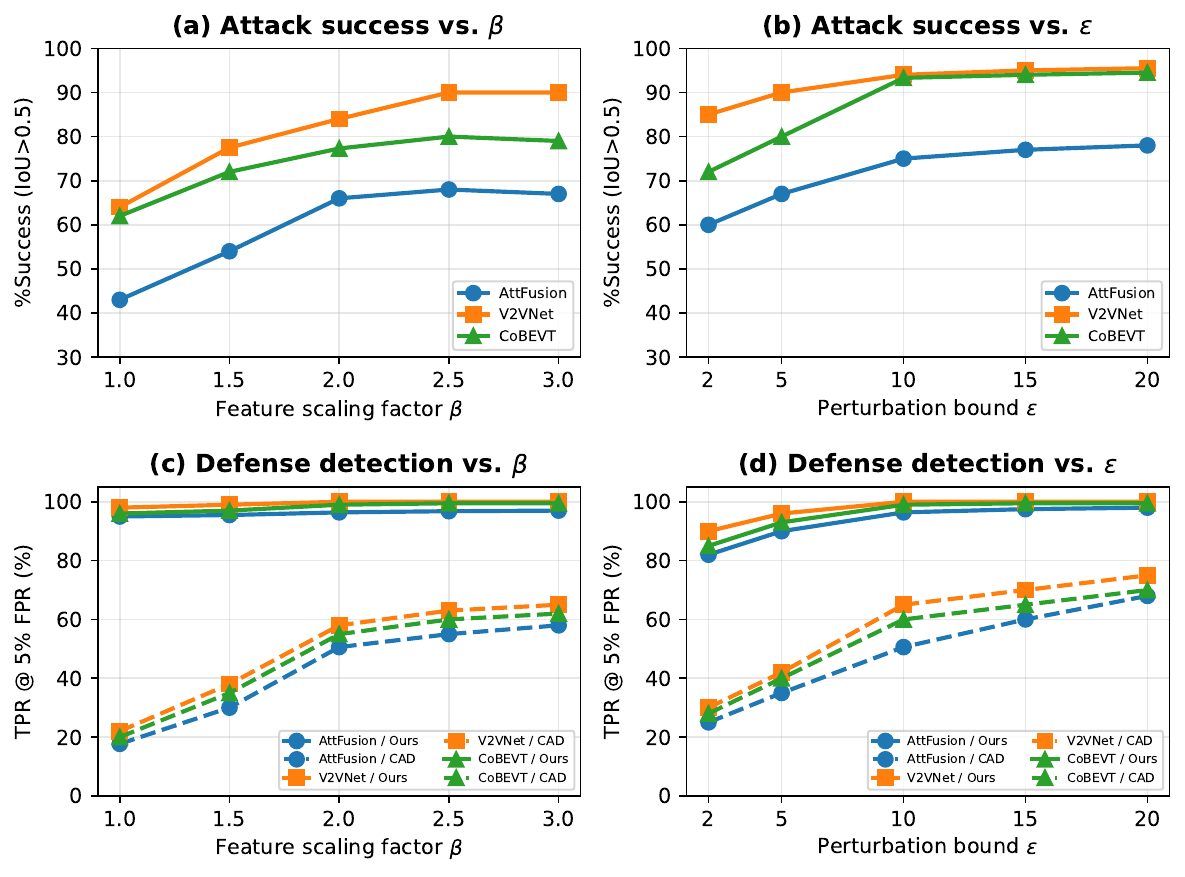}
\captionof{figure}{Parameter sensitivity: attack success and defense vs.\ $\beta$ and $\varepsilon$.}
\label{fig:factor_params}
\end{minipage}
\end{figure*}

Across all three OPV2V models (Table~\ref{tab:evaluation_perception_results}), the full attack (\modelName) achieves $>$94\% improvement rate and 75--94\% success rate at IoU$>$0.5. V2VNet is the most vulnerable (94.0\% success, average IoU=0.70), likely because its spatial-aware message passing amplifies manipulated features more strongly during fusion. CoBEVT shows similarly high vulnerability (93.3\%) despite its transformer-based fusion, indicating that attention mechanisms do not inherently resist feature-level manipulation.

\textbf{IoU shift is significant.}
Figure~\ref{fig:iou_distributions} visualizes the IoU distributions before and after attacks on AttFusion/OPV2V. In the normal case (blue), IoU with the target is concentrated near zero. After the full attack (red), a substantial mass shifts above 0.5, with V2VNet and CoBEVT showing the sharpest peaks near 0.7--0.8.

Each attack component contributes incrementally: ray-cast ($\beta=1$) achieves 42--68\% success; beta scaling raises it to 59--90\%; \modelName adds +15.7\% for AttFusion. The PGD baseline achieves only 9--15\% success despite $\sim$2$\times$ more computation per frame.

\subsubsection{Transfer Attack in Heterogeneous Collaboration}
\label{sec:eval_perception_transfer}

We evaluate cross-backbone transfer using HEAL~\cite{heal}, which aligns features from different architectures (m1--m4) into a shared space. The attacker trains \modelName on m1 and attacks a victim running m3 through HEAL's alignment layers, which is a realistic attack vector when the deployment involves mixed vehicle models and sensors.

\textbf{Cross-modality transfer is effective.}
In the heterogeneous setting (m1$\to$m3), the attack achieves 95.9\% improvement rate and 87.5\% success rate at IoU$>$0.5 (average IoU=0.64). This \emph{exceeds} the same-modality baseline (m1$\to$m1): 86.5\% improvement and 73.3\% success (IoU=0.58). The result confirms that our attack operates at the spatial coordination level---manipulating the pose relationship between attacker and victim---rather than exploiting architecture-specific features, making it inherently transferable across backbone types. HEAL's alignment mechanism, designed to harmonize heterogeneous features, does not mitigate the attack because the feature distortion originates from a physically shifted pose, which the alignment layer faithfully preserves.

\subsubsection{Defense Resilience}
\label{sec:eval_defense_resilience}

Table~\ref{tab:evaluation_perception_results} reports detection rates (TPR at 5\% FPR); Figures~\ref{fig:defense_dist} and~\ref{fig:defense_roc} show score distributions and ROC curves. For fair comparison, only \defenseName's stage~3 anomaly detection is applied (presuming the affected object is safety-critical).

\textbf{Global feature-level defenses are near-random.}
Global LUCIA achieves only 7--10\% TPR because the perturbation affects $<$0.04\% of the feature map, making the global L1 signal negligible. The trust score changes by only 0.015 on average (0.509$\to$0.494 for AttFusion), so attention reweighting has no meaningful mitigation effect. MADE performs similarly (7--9\% TPR), as the perturbation changes feature \emph{location}, not \emph{content}. ROBOSAC~\cite{li2023among} and MATE~\cite{hallyburton2025security} also fail ($\leq$5\% TPR): the shifted detection remains plausible, and per-agent trust scores stay stable.

\textbf{CAD provides partial, shift-dependent detection.}
CAD detects 47--73\% of full attacks on OPV2V when shifts are 0.5--2.0\,m (Table~\ref{tab:evaluation_perception_results}). However, CAD's effectiveness degrades sharply with shift size: at shifts $<$0.5\,m (the scenario-relevant regime), detection drops to 4.6\% because the spoofed detection largely overlaps with the original occupied zone (Figure~\ref{fig:defense_dist}c). This makes CAD ineffective against the small per-frame perturbations used in scenario attacks ($\epsilon_{\text{loc}}=0.5$\,m). On V2X-Real, CAD drops to 21.5\% (AUC\,=\,0.66) because the real-world occupancy maps---estimated via RANSAC and clustering without ground-truth map data---produce high baseline disagreement scores even on clean frames.

\textbf{\defenseName achieves robust detection regardless of shift distance.}
Our per-object raw L1 feature comparison detects 82--99\% of full attacks across all three OPV2V models at 5\% FPR. The key insight is that $\beta$ scaling amplifies the attacker's features by a factor of 2--3$\times$, creating a magnitude anomaly that is invisible at the global level ($<$0.04\% of voxels affected) but clearly separable when the comparison is localized to the target object's bounding box region. For the ray-cast-only variant ($\beta=1$), \defenseName detection drops to 9--15\% because the injected features have normal magnitude. Figure~\ref{fig:defense_dist}d also shows \defenseName distinguishes attacks at small shifts $<$0.5\,m. In Figure~\ref{fig:defense_roc}, \defenseName dominates at all operating points, achieving AUC$>$0.98 on the three OPV2V settings and 0.94 on V2X-Real. The lower V2X-Real AUC reflects the smaller $\beta=1.2$ (versus $2.0$--$3.0$ on OPV2V), which produces weaker feature amplification and more overlap with normal scores.





\begin{table*}[t]
\begin{minipage}[t]{0.70\textwidth}
\vspace{0pt}
  \small
  \centering
  \caption{Scenario attack effectiveness and defense detection rates.}
  \label{tab:evaluation_scenario_results}
  \setlength{\tabcolsep}{2pt}
  \begin{tabular}{| l | l | c | c | c | c | c | c | c | c | c |}
    \noalign{\global\arrayrulewidth1pt}\hline\noalign{\global\arrayrulewidth0.4pt}
    \multirow{2}{*}{\textbf{Setting}} & \multirow{2}{*}{\textbf{Optimize}} & \multicolumn{5}{c|}{\textbf{Attack Effectiveness}} & \multicolumn{4}{c|}{\textbf{Defense TPR@5\%FPR}} \\
    \cline{3-11}
    & & ADE ($\Delta$) & FDE ($\Delta$) & MinDist ($\Delta$) & \%Imp & \%Danger & CAD & LUCIA & MADE & Ours \\
    \noalign{\global\arrayrulewidth1pt}\hline\noalign{\global\arrayrulewidth0.4pt}
    \multirow{3}{*}{\shortstack[l]{AttFusion\\OPV2V}}
    & White-Box & \textbf{9.74}\ts{+9.08} & \textbf{17.85}\ts{+16.02} & 4.40\ts{$-$0.23} & 45.1 & 24.5 & 11.0 & 5.3 & 6.8 & \textbf{80.0} \\
    & Query-Access & 2.54\ts{+1.88} & 5.15\ts{+3.32} & 4.36\ts{$-$0.27} & \textbf{60.4} & 18.8 & 13.1 & 5.3 & 6.8 & \textbf{83.8} \\
    & Transfer & 5.11\ts{+3.19} & 10.41\ts{+4.78} & \textbf{3.93}\ts{$-$0.72} & 58.8 & \textbf{50.0} & 11.0 & 5.3 & 6.8 & \textbf{80.0} \\
    \hline
    \multirow{3}{*}{\shortstack[l]{V2VNet\\OPV2V}}
    & White-Box & \textbf{9.78}\ts{+9.12} & \textbf{17.70}\ts{+15.87} & 4.16\ts{$-$0.47} & 68.6 & \textbf{46.1} & 8.1 & 8.2 & 7.3 & \textbf{57.6} \\
    & Query-Access & 2.71\ts{+2.05} & 5.38\ts{+3.55} & 4.24\ts{$-$0.39} & 67.6 & 28.4 & 6.1 & 8.5 & 7.0 & \textbf{58.6} \\
    & Transfer & 5.29\ts{+2.96} & 10.08\ts{+3.27} & \textbf{3.73}\ts{$-$0.92} & \textbf{71.6} & 38.2 & 8.1 & 8.2 & 7.3 & \textbf{57.6} \\
    \hline
    \multirow{3}{*}{\shortstack[l]{CoBEVT\\OPV2V}}
    & White-Box & \textbf{9.79}\ts{+9.13} & \textbf{18.94}\ts{+17.11} & 4.39\ts{$-$0.24} & 62.7 & 38.2 & 5.0 & 4.8 & 7.0 & \textbf{77.0} \\
    & Query-Access & 2.38\ts{+1.72} & 4.77\ts{+2.94} & 4.39\ts{$-$0.24} & \textbf{69.6} & 23.5 & 6.2 & 4.8 & 7.5 & \textbf{76.0} \\
    & Transfer & 5.25\ts{+3.08} & 10.10\ts{+3.46} & \textbf{4.25}\ts{$-$0.49} & 71.4 & \textbf{45.9} & 5.0 & 4.8 & 7.0 & \textbf{77.0} \\
    \hline
    \multirow{3}{*}{\shortstack[l]{AttFusion\\V2X-Real}}
    & White-Box & \textbf{5.68}\ts{+2.61} & \textbf{10.93}\ts{+4.90} & \textbf{6.41}\ts{$-$0.72} & 46.7 & \textbf{11.1} & 5.5 & 4.4 & 5.5 & \textbf{48.9} \\
    & Query-Access & 5.59\ts{+2.50} & 10.97\ts{+4.90} & 6.58\ts{$-$0.42} & \textbf{57.8} & 6.7 & 6.5 & 4.4 & 5.5 & \textbf{53.3} \\
    & Transfer & 5.20\ts{+1.98} & 10.16\ts{+3.80} & 6.25\ts{$-$0.61} & 54.5 & 9.1 & 5.5 & 4.4 & 5.5 & \textbf{51.1} \\
    \noalign{\global\arrayrulewidth1pt}\hline\noalign{\global\arrayrulewidth0.4pt}
  \end{tabular}

\end{minipage}%
\hfill
\begin{minipage}[t]{0.28\textwidth}
\vspace{0pt}
\small
\centering
\captionof{table}{Computational overhead per frame.}
\label{tab:overhead}
\setlength{\tabcolsep}{3pt}
\begin{tabular}{l r l}
    \hline
    \textbf{Component} & \textbf{Latency} & \textbf{Notes} \\
    \hline
    Ray casting & 20\,ms & 4 LiDARs \\
    Encoding & \multirow{2}{*}{5\,ms} & \\
    + Scatter  &                         & \\
    \modelName & $<$1\,ms & 40K params \\
    \textbf{Total attack} & \textbf{36\,ms} & Real-time \\
    \hline
    PGD baseline & 70\,ms & 1 grad step \\
    \hline
    Scenario (QA) & 20\,ms & 12 queries \\
    Scenario (WB) & 5\,ms & \\
    \hline
    \defenseName & $<$5\,ms & L1 crops \\
    \hline
\end{tabular}
\end{minipage}
\end{table*}

\subsubsection{Impacting Factors and Parameters}
\label{sec:eval_factors}


Figure~\ref{fig:factor_distance} analyzes four factors that affect attack success.

\textbf{The attack is robust to geometry.}
Attack success depends on the geometric influence of benign vehicles: their distance, number, and point density around the target. When benign vehicles are close ($<10$\,m), their features strongly oppose the attacker during fusion, reducing IoU (e.g., 0.63$\rightarrow$0.60), whereas beyond 20\,m their influence diminishes, allowing the attacker to dominate. Increasing the number of benign vehicles (1--3) slightly reduces IoU but does not eliminate the attack, as $\beta$-scaled attacker features still outweigh individual contributions. Similarly, higher benign point density near the target ($>$100 points within 2\,m) strengthens competing features, reducing IoU from 0.60 to 0.57. Overall, attack effectiveness decreases with stronger benign geometric support but remains robust under typical conditions.

\textbf{Number of colluding attackers matters but is not a game changer.} Figure~\ref{fig:factor_distance}d shows attack effectiveness when multiple attackers target the same object independently and simultaneously.
Adding attackers boosts average IoU from 0.53 (1 attacker) to 0.62 (2 attackers) and 0.64 (3 attackers). The benefit saturates beyond 2 attackers, as 2 colluding $\beta$-scaled feature maps already dominate fusion against 1--2 benign vehicles.


\textbf{Attack effectiveness \& stealthiness trade-off.}
Figure~\ref{fig:factor_params} examines sensitivity to key attack parameters. The scaling factor $\beta$ (Figure~\ref{fig:factor_params}a) exhibits a tradeoff: higher $\beta$ amplifies the attacker's features to dominate fusion but also distorts the feature structure, reducing detection quality. The optimal $\beta$ varies by model: AttFusion peaks at 2.0--2.5, while V2VNet continues improving up to 3.0 due to its more robust fusion mechanism. The perturbation bound $\varepsilon$ (Figure~\ref{fig:factor_params}b) shows monotonic improvement, with diminishing returns beyond $\varepsilon=10$.
From a defense perspective (Figure~\ref{fig:factor_params}c,d), \defenseName maintains $>$95\% detection across all $\beta$ values above 1.5 and all $\varepsilon$ values above 5, demonstrating robustness to parameter choices. CAD detection increases gradually from 18\% at $\beta=1.0$ to 58\% at $\beta=3.0$, tracking the increase in spoofed area but never approaching \defenseName's detection rate.

\subsubsection{Ablation Study}
\label{sec:eval_ablation}

\begin{figure*}[t]
\begin{minipage}[t]{0.40\textwidth}
\vspace{0pt}
\centering
\includegraphics[width=\textwidth]{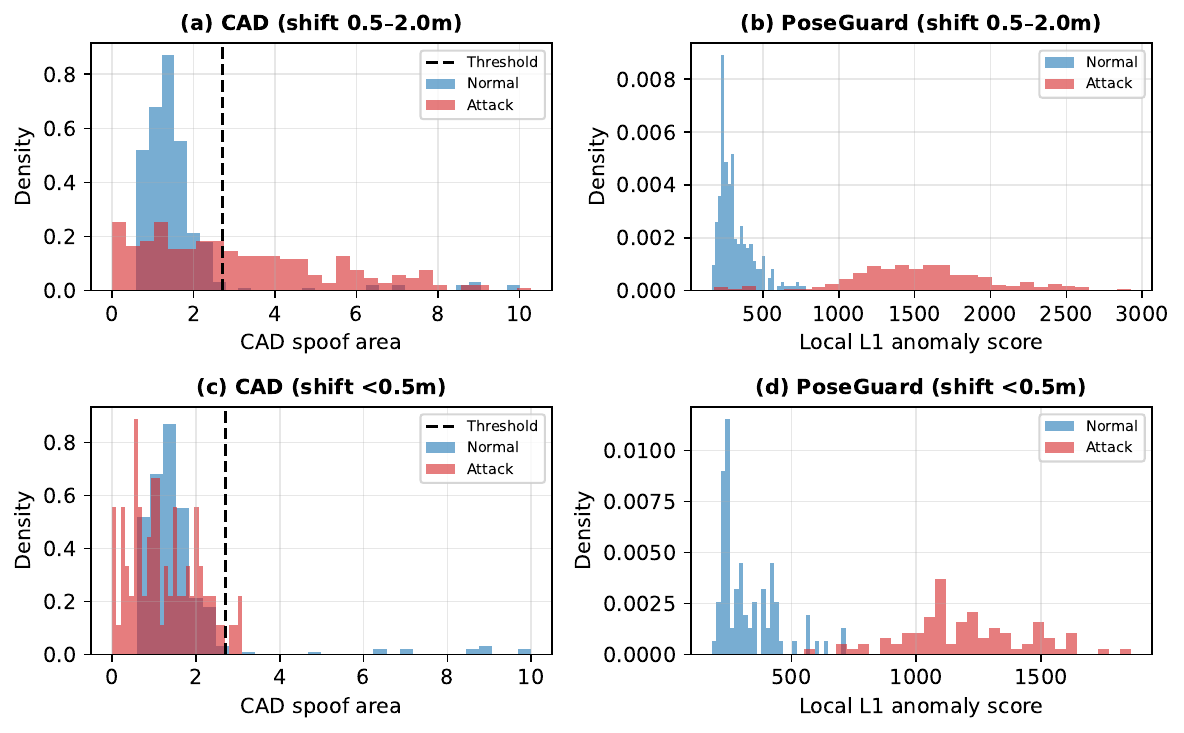}
\captionof{figure}{Defense score distributions.}
\label{fig:defense_dist}
\end{minipage}%
\hfill
\begin{minipage}[t]{0.16\textwidth}
\vspace{0pt}
\centering
\includegraphics[width=0.98\textwidth]{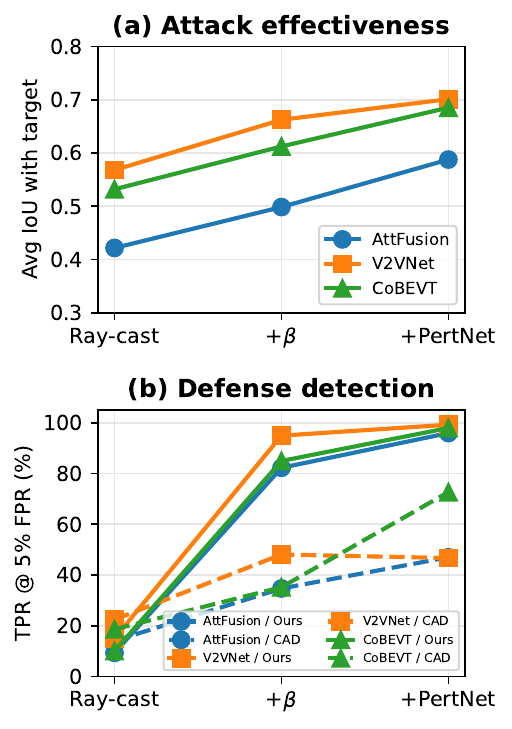}
\captionof{figure}{Ablation on attack variants.}
\label{fig:ablation}
\end{minipage}%
\hfill
\begin{minipage}[t]{0.40\textwidth}
\vspace{0pt}
\centering
\includegraphics[width=\textwidth]{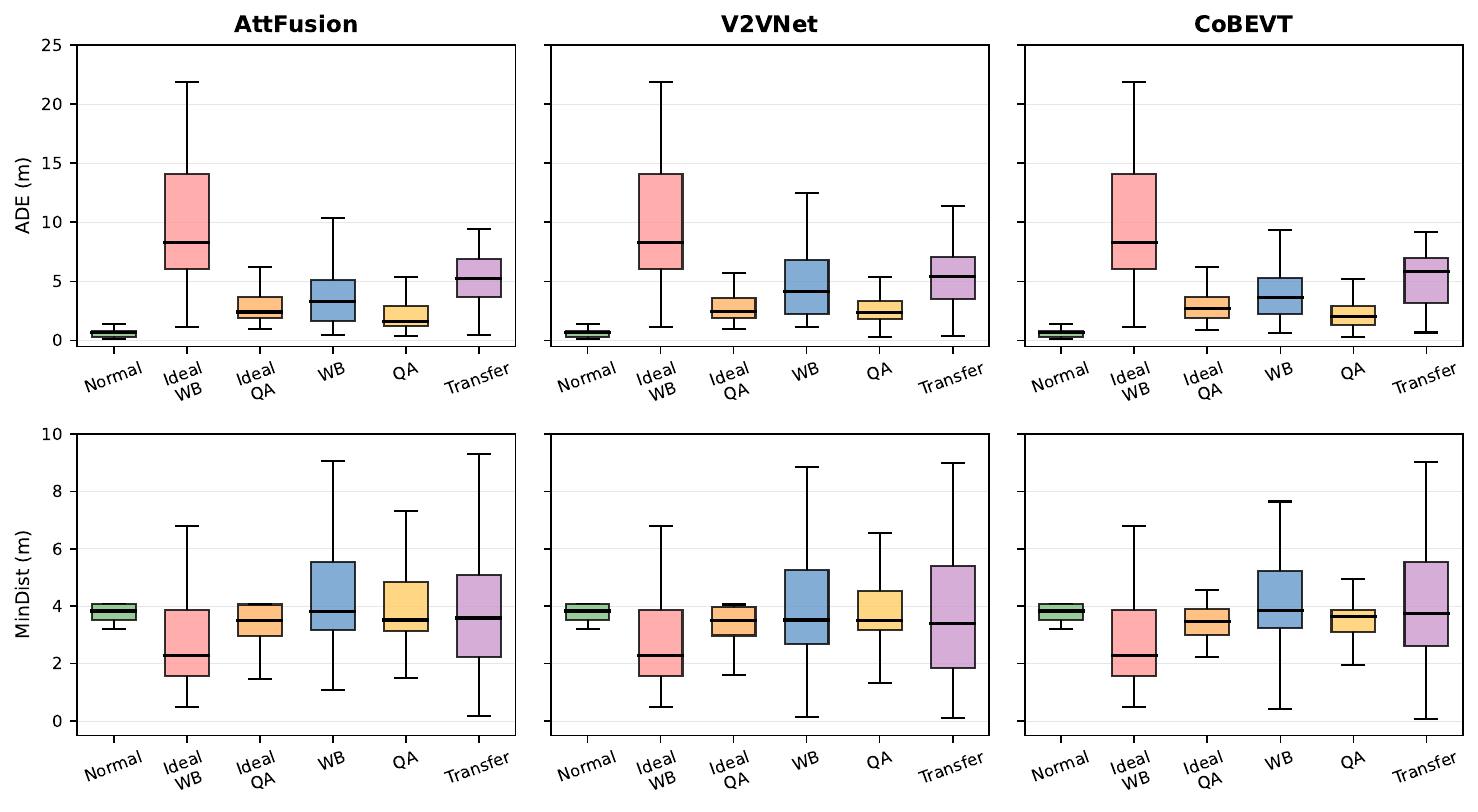}
\captionof{figure}{Scenario attack ADE and MinDist distributions.}
\label{fig:scenario_analysis}
\end{minipage}
\end{figure*}

We ablate the attack components to understand their individual contributions: (1) \textbf{Ray-cast only} ($\beta = 1$, no \modelName): Lower bound with naive initialization. (2) \textbf{Ray-cast + scaling} (best $\beta$, no \modelName): Adds feature amplification. (3) \textbf{Full attack} (best $\beta$ + \modelName): Adds learned perturbation correction. Figure~\ref{fig:ablation} shows the impact of components on both the attack and defense results.


\textbf{Each component is helpful for the attack.} In (a), each attack component adds incremental value: feature scaling ($\beta$) provides the largest IoU improvement (0.42$\to$0.50 for AttFusion), while \modelName adds a further 0.09 by learning corrections conditioned on local feature context. V2VNet and CoBEVT follow the same trend with higher absolute IoU due to their greater vulnerability to feature-level manipulation.

\textbf{Components also reveal an attacker--defender tradeoff.}
In (b), \defenseName's detection rate jumps sharply from 9\% (ray-cast only) to 82\% (+$\beta=2.0$) to 96\% (full attack), because $\beta$-scaling is the primary detectable signal---it amplifies the attacker's features by 2--3$\times$ within the target region, creating a magnitude anomaly that is invisible globally but clearly separable per-object. Without $\beta$ scaling, the ray-cast features have normal magnitude and are essentially undetectable by L1 comparison. CAD's detection rate increases gradually (14\%$\to$35\%$\to$50\%) mainly because the attack is becoming more successful thus exposes more spatial inconsistency, but lagging behind \defenseName at every operating point.

\subsection{Scenario Attack Results}
\label{sec:eval_scenario}

We evaluate the end-to-end safety impact of \attackName through the full autonomous driving stack. The scenario attacker optimizes an adversarial trajectory over $K{=}3$ frames (\S\ref{sec:attack_strategy}) under three modes:
\begin{itemize}
    \item \textbf{White-Box}~(WB), where the attacker computes exact gradients through the victim's GRIP++ predictor.
    \item \textbf{Query-Access}~(QA), where it estimates gradients via finite-difference queries.
    \item \textbf{Transfer}, where it optimizes on GRIP++ but the victim uses Trajectron++, testing cross-model generalization.
\end{itemize}

\subsubsection{Effectiveness}

We evaluate scenario attack using five metrics:
\begin{itemize}
    \item \textbf{ADE/FDE~($\Delta$)}, the average/final displacement error of the victim's predicted target trajectory relative to ground truth (delta from normal prediction).
    \item \textbf{MinDist~($\Delta$)}, the minimum predicted distance between the target and victim over the prediction horizon (lower = more dangerous).
    \item \textbf{\%Improved}, the fraction of cases where MinDist decreases.
    \item \textbf{\%Danger}, the fraction of cases where MinDist$<$3\,m, indicating the predicted target trajectory closely approaches the victim and would trigger emergency braking in AV software~\cite{apollo,autoware}.
\end{itemize}

Table~\ref{tab:evaluation_scenario_results} and Figure~\ref{fig:scenario_analysis} summarize the results. The white-box attack produces large ADE ($>$9\,m) via direct gradient optimization, with danger rates of 25--46\% across models. The query-access attack achieves moderate deflection (ADE 2--3\,m) with 19--28\% danger. The transfer attack (GRIP++$\to$Trajectron++) achieves 38--50\% danger, comparable to or exceeding white-box, because Trajectron++ propagates observation noise aggressively into long-horizon predictions.

\textbf{Ideal vs.\ actual gap.} Figure~\ref{fig:scenario_analysis} analyzes prediction error distributions. The Ideal white-box target (median ADE$\approx$9\,m) is substantially larger than the achieved white-box ADE (median$\approx$3\,m), reflecting the gap between the optimizer's desired trajectory deflection and what the perception attack actually achieves. The perception attack is the bottleneck: a 0.5\,m per-frame shift produces only $\sim$1.4\,m of cumulative tracking error, which is then amplified to $\sim$3--5\,m ADE by the predictor. Yet this constraint is a desired consequence of stealthiness (\S\ref{sec:eval_defense_resilience}). For MinDist (bottom row), Normal cases cluster tightly around 4.6\,m. The Ideal white-box target reduces this to $\sim$2.5\,m, and actual attacks achieve median MinDist of 3.7--4.4\,m depending on the model and variant. V2VNet shows the largest MinDist reduction (median 3.7\,m for Transfer), consistent with its higher perception-level vulnerability.

\textbf{White-box optimization is significantly stronger.} The Ideal query-access target (median ADE$\approx$2--3\,m) is more conservative than Ideal white-box ($\approx$9\,m), because finite-difference gradients are noisier, naturally limiting the optimizer's ambition. The actual query-access outcomes closely match the ideal, suggesting the optimizer sets achievable targets.

\textbf{Transfer attack works across predictors.} Transfer attacks achieve comparable median ADE (4--5\,m vs.\ 3--5\,m) and MinDist distributions to white-box despite using a different predictor, confirming that the attack operates primarily through tracking manipulation: the corrupted tracker state produces unsafe predictions regardless of which predictor processes it.

\subsubsection{Defense Resilience}

Defense detection rates are in Table~\ref{tab:evaluation_scenario_results} (transfer rates equal white-box since only the predictor differs). The per-frame attack uses the same $\beta$-scaling and \modelName, so feature-level anomaly behavior matches \S\ref{sec:eval_effectiveness}. However, per-frame shift is bounded by $\epsilon_{\text{loc}}=0.5$\,m (vs.\ 0.5--2.0\,m for perception cases), which affects CAD but not \defenseName. The affected object is close to the victim ($<$10\,m), so it is reliably selected as safety-critical.

\textbf{Defense effectiveness holds for scenario attacks}. \defenseName achieves 58--84\% TPR at 5\% FPR across all OPV2V settings---lower than the 82--99\% for perception attacks (Table~\ref{tab:evaluation_perception_results}) because the 0.5\,m per-frame shift is sub-voxel in some cases (voxel size 0.4\,m), producing minimal feature displacement even with $\beta$-scaling. Nevertheless, \defenseName remains the most effective defense by a large margin.
CAD detection drops to 5--13\% for scenario attacks, compared to 14--73\% for perception attacks, because the small per-frame shift keeps the spoofed detection within the original occupied zone. LUCIA and MADE remain at 5--8\% (near-random), consistent with their perception-level performance.

\begin{figure}[t]
\centering
\includegraphics[width=0.49\textwidth]{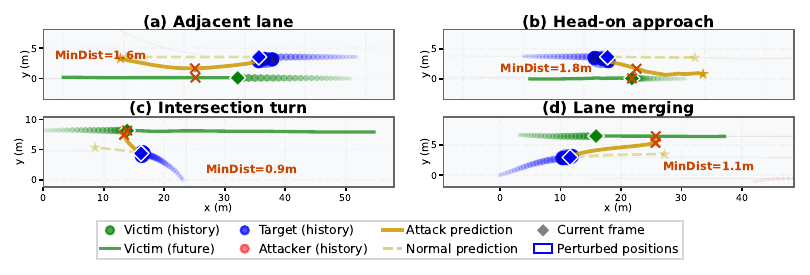}
\caption{Case studies of move-in scenario attacks on AttFusion/OPV2V, covering four types of traffic interactions.}
\label{fig:case_study}
\end{figure}

\subsubsection{Case Study}
\label{sec:eval_case_study}

Figure~\ref{fig:case_study} shows four representative white-box scenario attacks on AttFusion/OPV2V. More cases: Appendix~\ref{sec:appendix_case_studies}.

The details are below. \textbf{Adjacent lane} (a): target and victim travel in the same direction in adjacent lanes; the attack shifts the target's perceived position into the victim's lane, inducing braking for a phantom cut-in (minDist=2.16\,m, ADE=2.40\,m).
\textbf{Head-on approach} (b): the target travels toward the victim from the opposite direction; three small shifts make the tracker predict a trajectory crossing into the victim's path (minDist=1.77\,m, ADE=5.05\,m).
\textbf{Intersection turn} (c): the target approaches at $\sim$145$^\circ$ to the victim at an intersection; lateral shifts redirect the predicted trajectory toward the victim, producing the closest approach across all four scenarios (minDist=0.87\,m, ADE=1.70\,m).
\textbf{Lane merging} (d): the target merges from an adjacent curved lane with the attacker in opposing traffic; the attack deflects the prediction toward the victim (minDist=2.32\,m, ADE=2.78\,m).

\subsection{Closed-Loop Attack Simulation}
\label{sec:eval_closed_loop}

The dataset evaluation replays logged trajectories without victim reaction. To verify that corrupted predictions change actual driving behavior, we reproduce attacks in closed-loop CARLA simulation where the victim perceives, plans, and actuates every frame.

We build on CARLA~0.9.16~\cite{carla} (Town10HD, synchronous 20\,Hz) with two collaborative vehicles: a victim (ego) running the full AV stack---AttFusion~\cite{xu2022opv2v} perception with PointPillars~\cite{lang2019pointpillars} backbone, AB3DMOT~\cite{weng2020ab3dmot} tracking, GRIP++~\cite{li2019grip++} prediction, and adaptive cruise control---and one collaborator that doubles as the attacker, executing \attackName online (Algorithm~\ref{alg:online_attack}, $\beta=2$, \modelName) under the same $\epsilon_{\text{loc}}=0.5$\,m stealth bound. Each agent carries a LiDAR matching the OPV2V spec, so the OPV2V-trained model transfers without retraining. We evaluate two scenarios across randomized road configurations each, covering both attack objectives from \S\ref{sec:scenario_challenges}.

\begin{figure}[t]
\centering
\begin{subfigure}{0.23\textwidth}
\includegraphics[width=\textwidth]{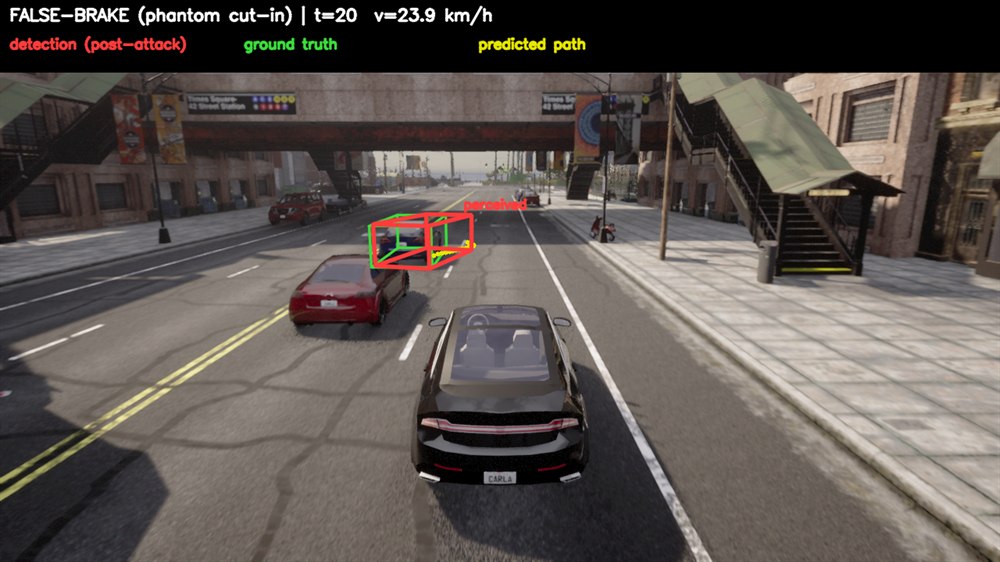}
\caption{Move-in attack induces a fake lane changing behavior.}
\end{subfigure}
\hfill
\begin{subfigure}{0.23\textwidth}
\includegraphics[width=\textwidth]{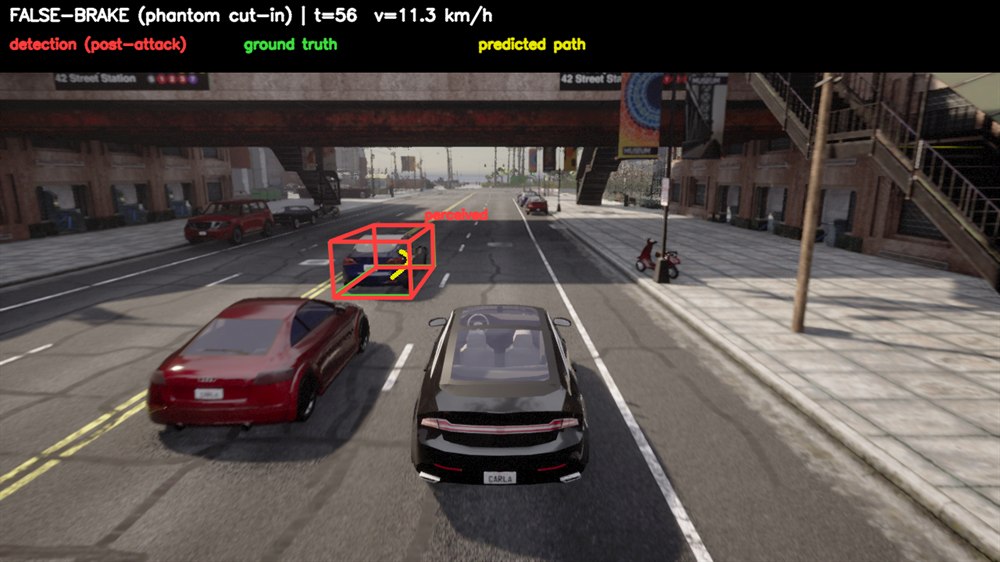}
\caption{Move-in attack triggers the hard braking at clear road.}
\end{subfigure}

\begin{subfigure}{0.23\textwidth}
\includegraphics[width=\textwidth]{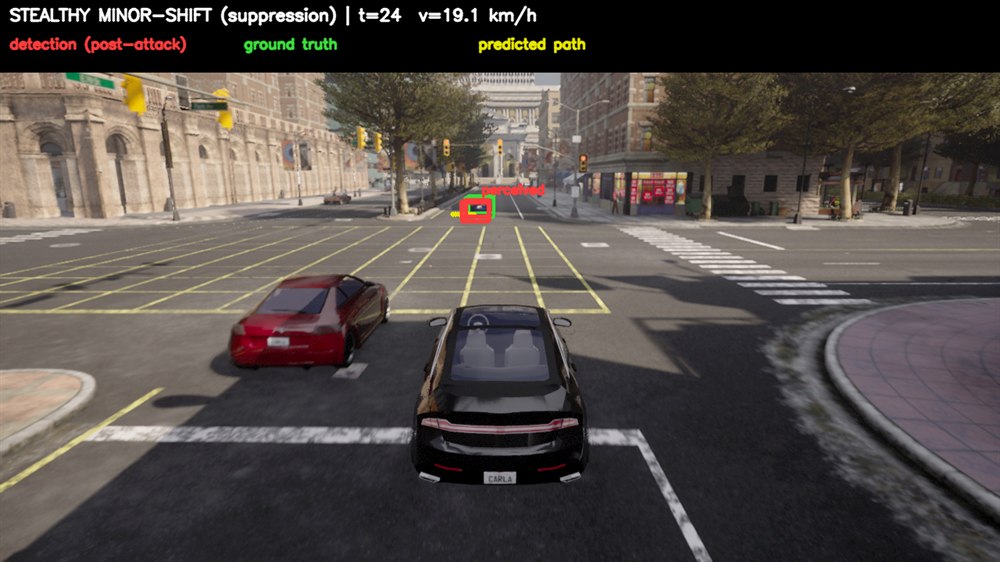}
\caption{Move-out attack pushes an obstacle away from ego lane.}
\end{subfigure}
\hfill
\begin{subfigure}{0.23\textwidth}
\includegraphics[width=\textwidth]{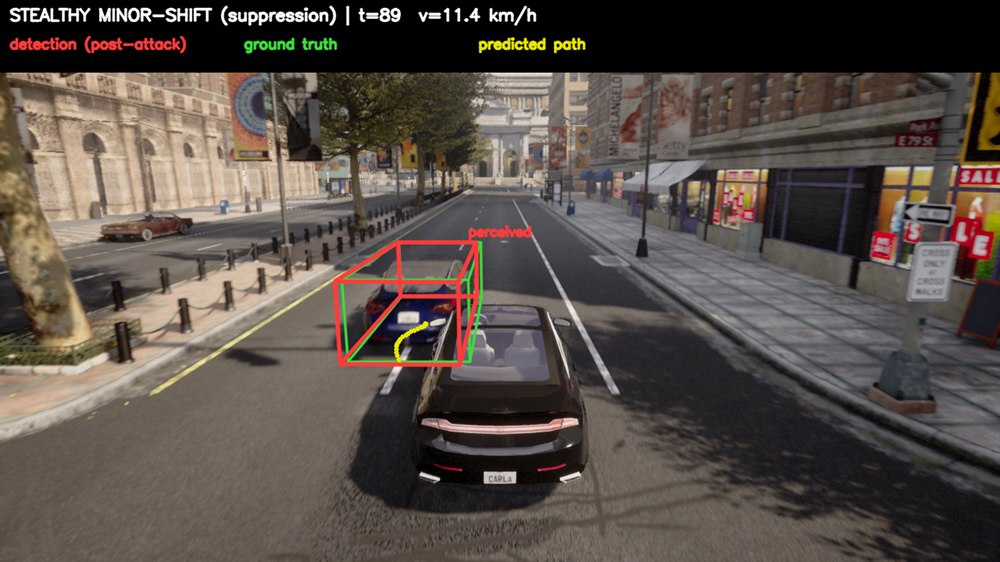}
\caption{Move-out attack triggers a collision.}
\end{subfigure}
\caption{Closed-loop attack simulation in CARLA. Illustrating a hard braking case (top) and a collision case (bottom).}
\label{fig:carla}
\end{figure}

\textbf{Move-in attacks} (false braking): a vehicle in an adjacent lane is shifted into the ego lane, triggering false braking on a clear road (Figure~\ref{fig:carla}, top).
The attack induces unwarranted braking on 9/12 (75\%) of cruise-clean roads. The attack slows the victim car by 11.7 km/h on average, a \textbf{52\%} speed reduction in a second. In the worst case, the victim brakes itself from $\ge 22$\,km/h down to 6.5\,km/h (70\% speed reduction) on a fully clear road.

\textbf{Move-out attacks} (collision risks): a stopped vehicle marginally blocking the ego lane is shifted out of lane, suppressing braking and causing collision (Figure~\ref{fig:carla}, bottom).
In 4/10 (40\%) cases the decision of braking is delayed (the time-to-collision drops from 2.6s to below 1.5s), introducing significantly increased collision risks.
In the worst case, the vehicle keeps driving at 22\,km/h and cannot stop safely with only a 0.1\,s time-to-collision.

Our defense \defenseName also detects the attack in the above scenarios (in 22/22 runs, 100\%), regardless of whether the attacks are successful. These results confirm that the pose perturbation measured on logged data propagates end-to-end through a reactive driving stack. For more cases please refer to Appendix~\ref{sec:appendix_carla_cases}.

\subsection{Computational Overhead}
\label{sec:eval_overhead}

Table~\ref{tab:overhead} summarizes per-component latency (on a server with NVIDIA L40S GPU and AMD EPYC 9354 CPU). The full perception attack ($\sim$36\,ms) is real-time capable and 2$\times$ faster than PGD-based methods ($>$70\,ms). \defenseName adds $<$5\,ms. The scenario optimizer adds only 5\,ms in white-box mode, as planning runs in parallel with perception attack execution.

\section{Discussion}
\label{sec:discussion}

\myparagraph{Defense alternatives}
Beyond detection-based defenses (\S\ref{sec:eval_defense_resilience}), two architectural alternatives exist. First, the victim could discard remote data when local sensing is sufficient, but reliably doing so without losing coverage for occluded objects remains open~\cite{hallyburton2025probabilistic}. Second, TEEs could attest pipeline integrity~\cite{hu2020cvshield}, but they require dedicated hardware and introduce attestation latency unsuitable for real-time communication.

\myparagraph{Limitations}
(1)~We evaluate academic open-source implementations rather than closed-source industry systems with possible undisclosed defenses.
(2)~The attack requires a vulnerable traffic scenario with a suitable target near the victim.
(3)~It is probabilistic due to limited downstream knowledge and environmental uncertainty.
(4)~It assumes a modular AV stack; end-to-end driving models are outside our scope.
(5)~\defenseName reduces the attack budget and success rate but is not an assured cure.

\section{Conclusion}

We present a stealthy, scenario-aware data fabrication attack on collaborative perception. It combines multi-view ray casting with a learned network to manipulate object poses in real time while evading anomaly detection, and orchestrates temporal attack sequences through an observe--predict--plan--execute loop to induce unsafe behaviors. We also propose an object-level defense that better detects localized perturbations, showing the need for end-to-end security evaluation beyond perception-level defenses.

\section*{Acknowledgment}
This work was supported by the National Science Foundation under Award CNS-2321532.

\bibliographystyle{plain}
\bibliography{cite}

\appendix 

\section{Open Science}

The artifact is at \url{https://github.com/zqzqz/PosePert}. It contains the
source for the perception attack, the scenario-aware strategy and \defenseName, together
with the per-model \modelName checkpoints (40K parameters each), our integration code and
weights for GRIP++~\cite{li2019grip++} and Trajectron++~\cite{salzmann2020trajectron++},
the pre-computed test cases (300 perception and 102 scenario per model), the cached
ray-cast point clouds and per-case results behind every table, and the figure generation
scripts, so the figures can be reproduced without re-running the experiments. The
perception models (AttFusion, V2VNet, CoBEVT) come from OpenCOOD~\cite{xu2022opv2v} and
the pre-trained weights are included. Both datasets are public: OPV2V~\cite{xu2022opv2v} and V2X-Real~\cite{xiang2024v2x}; no proprietary data is used.

Releasing attack code is dual-use. We mitigate this by targeting only simulated and
research datasets rather than deployed systems, releasing \defenseName alongside the
attack, and coordinating with the collaborative perception community before public
release.

\section{Ethical Considerations}

All experiments run on simulated (OPV2V) and controlled research (V2X-Real) datasets. No
attack was performed against a deployed autonomous driving system, public infrastructure
or real traffic participant. The attack also requires the adversary to be a registered
participant in the collaborative perception network, which confines the threat to insiders
rather than external exploitation.

The vulnerability is inherent to intermediate-fusion architectures rather than specific to
any deployed product, and we study widely used open-source research systems, so we will
coordinate with the OpenCOOD maintainers on defensive measures before any public code
release. We pair the attack with \defenseName, which reaches 58--99\% detection, and our
analysis of the attacker--defender tradeoff---higher $\beta$ raises attack success but also
detectability---is intended to inform defensive design. Understanding these threats across
the full pipeline, from perception through tracking and prediction to planning, is
necessary before connected autonomous vehicles are deployed at scale.

\section{Experimental Details}
\label{sec:appendix_details}

Setup details omitted from \S\ref{sec:eval_setup}. The implementation is $\sim$4{,}400
lines of Python: $\sim$1{,}700 perception attack, $\sim$1{,}400 scenario attack,
$\sim$1{,}300 defenses.

\myparagraph{Test case selection}
Perception cases enumerate every (attacker, victim, target) triple whose target is visible
to the attacker ($\geq$50 LiDAR points) and correctly detected by the unattacked model
(IoU $>$ 0.2), so post-attack error is attributable to the attack rather than to a
pre-existing failure. Survivors are stratified into 100 easy, 100 moderate and 100 hard by
target point count and range, otherwise the sample skews toward near, densely sampled
targets. The 300 cases span 136 scenes at 3.5--38\,m (mean 16.4\,m); \modelName's 1{,}000
training samples come from the train split, so no scene is shared. Scenario cases take
scenarios with $\geq$60 frames and $\geq$2 vehicles, pick the closest visible vehicle as
target, and keep those whose target-to-victim distance at \emph{frame~30}---about 1\,s
after the last attack frame, when the corrupted prediction acts---is under 8\,m. Targets
below 3\,m/s are dropped as degenerate, and each scenario is capped at 15 cases. The 102
cases cover 13 scenarios at 3.4--7.7\,m (mean 4.8\,m).

\myparagraph{Models and attack}
PointPillars discretizes the cloud into a $704 \times 200$ BEV grid at 0.4\,m with 64
channels per voxel; the backbone emits 384-channel fused features. That 0.4\,m cell sets
the attack's granularity: a shift below half a cell stays within one pillar. Ray casting
uses Open3D~\cite{zhou2018open3d} with four phantom LiDARs at 10\,m, each 64 beams
$\times$ 1{,}029 azimuth rays (0.35$^\circ$), giving 3{,}000--6{,}000 points per target
over a RANSAC ground plane. $\beta$ is grid-searched over $[1.0, 3.0]$ by \%Success
(IoU$>$0.5) on 50 held-out cases. \modelName is a 4-layer CNN ($\sim$40K parameters), 20--35
epochs of Adam (lr=$10^{-3}$) under an IoU loss through the frozen model. Blackbox
gradients use finite differences ($\delta{=}0.05$, 12 queries per iteration), so the
attacker needs only predicted trajectories; the Trajectron++~\cite{salzmann2020trajectron++}
baseline observes 8 frames and predicts 6 against GRIP++'s 20 and 20.

\myparagraph{Defenses}
Thresholds sit at the 95th percentile of clean scores, a 5\% false positive rate on benign
traffic. CAD~\cite{zhang2023data} builds occupancy maps from
SqueezeSegV3~\cite{xu2020squeezesegv3} segmentation and flags 1.7\,m$^2$ of spoofed area.
Global LUCIA~\cite{wang2025threat} uses full-feature-map L1 distance to the ego; global
MADE~\cite{zhao2024made} uses autoencoder reconstruction loss on $f_{\text{fused}} -
f_{\text{fused-without-agent}}$ (trained on 200 clean cases, 384 channels, 50 epochs). The
local variants restrict both to detected bounding boxes, because a pose perturbation
changes a small fraction of the map that a global statistic averages away.

\begin{figure}[t]
\centering
\begin{subfigure}{0.15\textwidth}
\includegraphics[width=\textwidth]{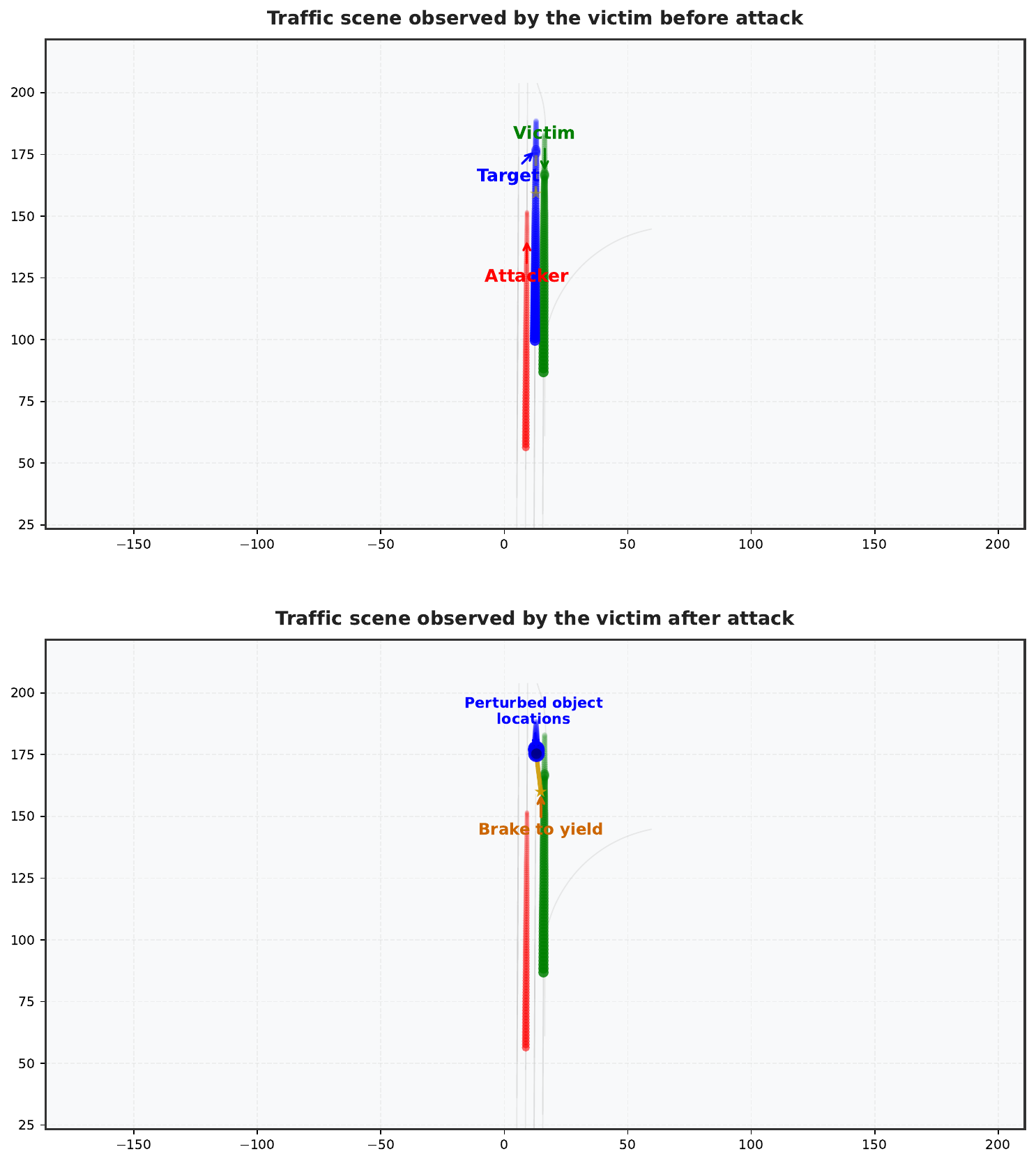}
\caption{Following vehicle}
\end{subfigure}
\hfill
\begin{subfigure}{0.15\textwidth}
\includegraphics[width=\textwidth]{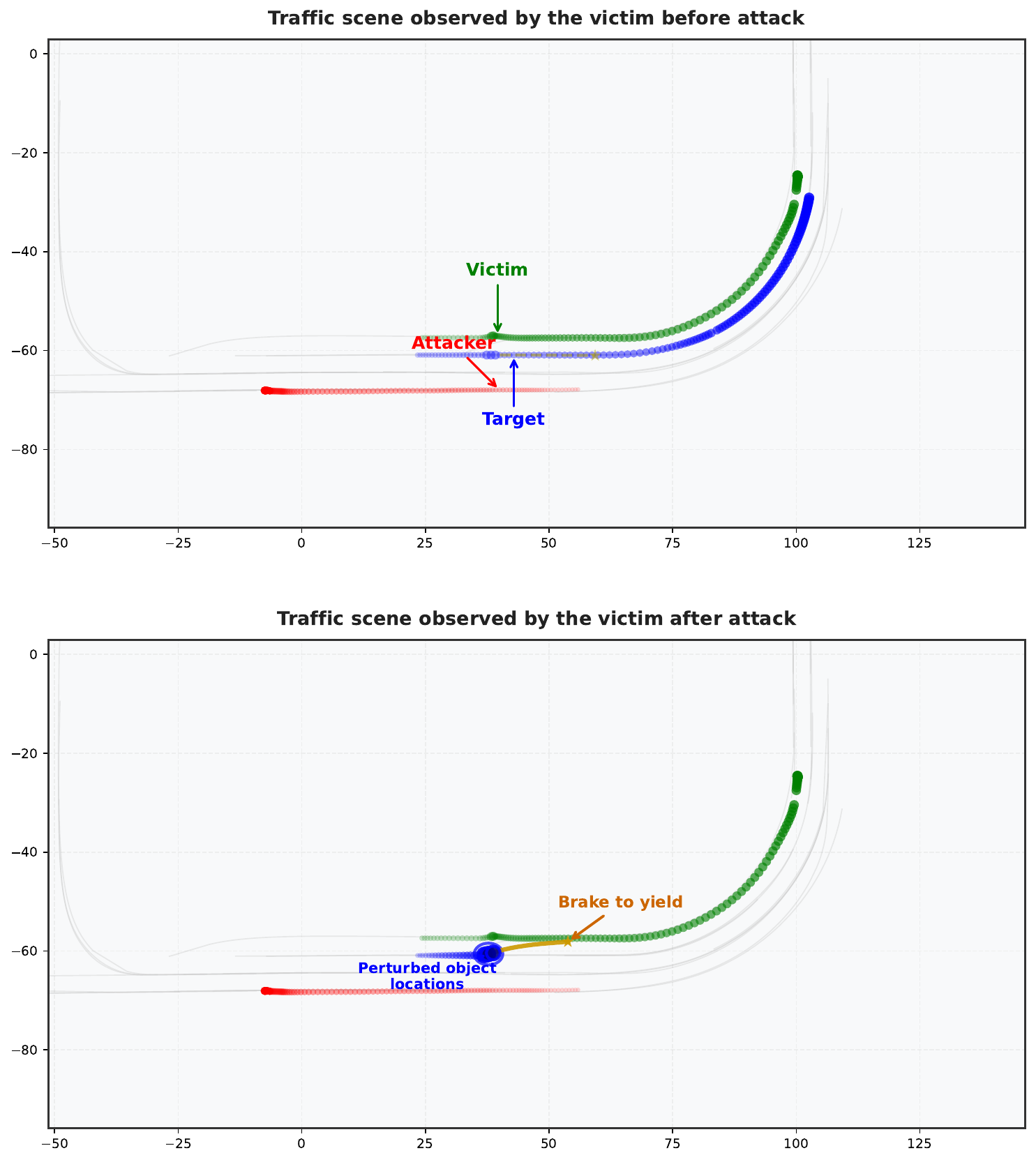}
\caption{Curved road}
\end{subfigure}
\hfill
\begin{subfigure}{0.15\textwidth}
\includegraphics[width=\textwidth]{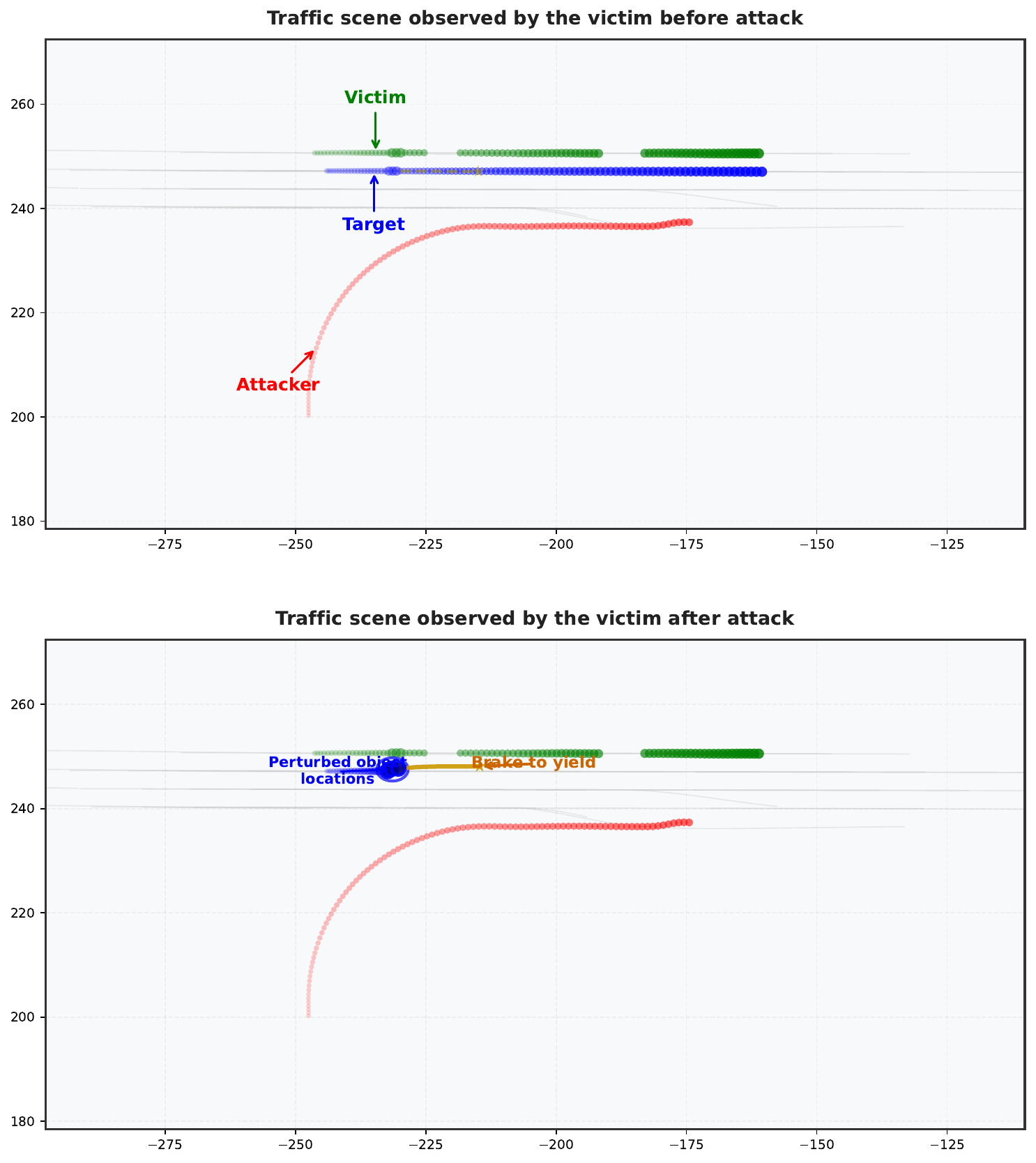}
\caption{Turning attacker}
\end{subfigure}
\\[2pt]
\begin{subfigure}{0.15\textwidth}
\includegraphics[width=\textwidth]{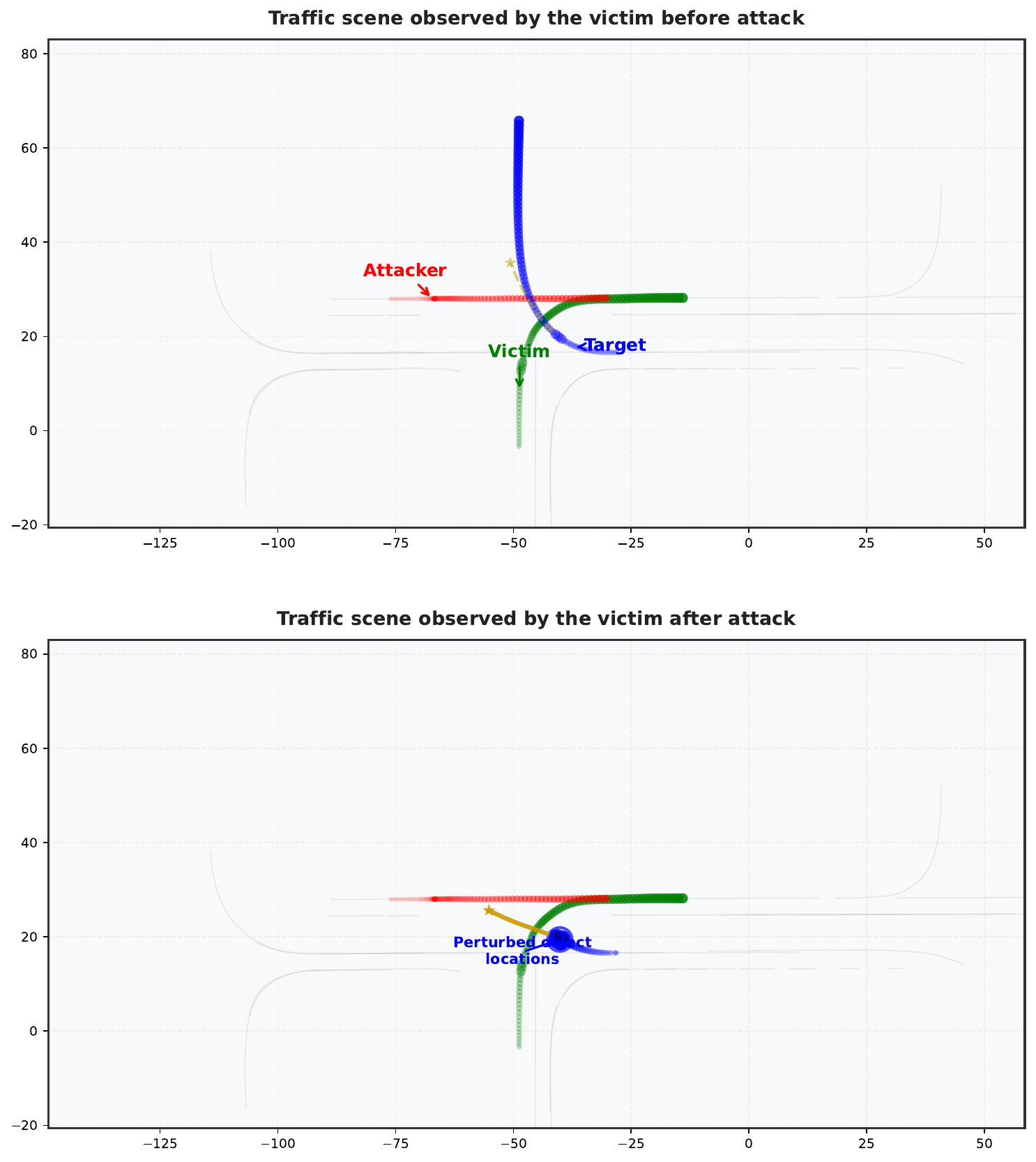}
\caption{Perpendicular crossing (fail)}
\end{subfigure}
\hfill
\begin{subfigure}{0.15\textwidth}
\includegraphics[width=\textwidth]{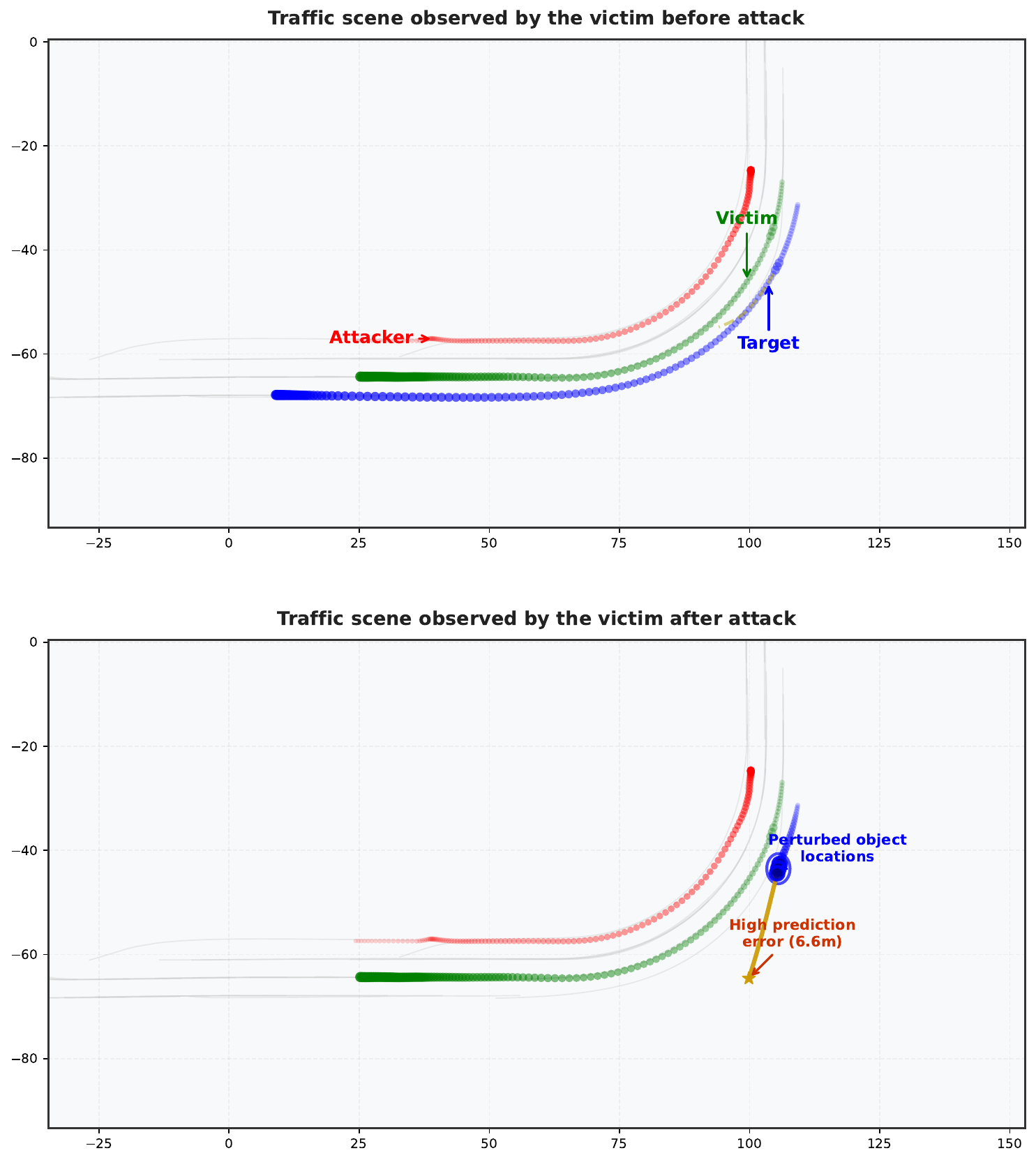}
\caption{High ADE, wrong direction (fail)}
\end{subfigure}
\hfill
\begin{subfigure}{0.15\textwidth}
\includegraphics[width=\textwidth]{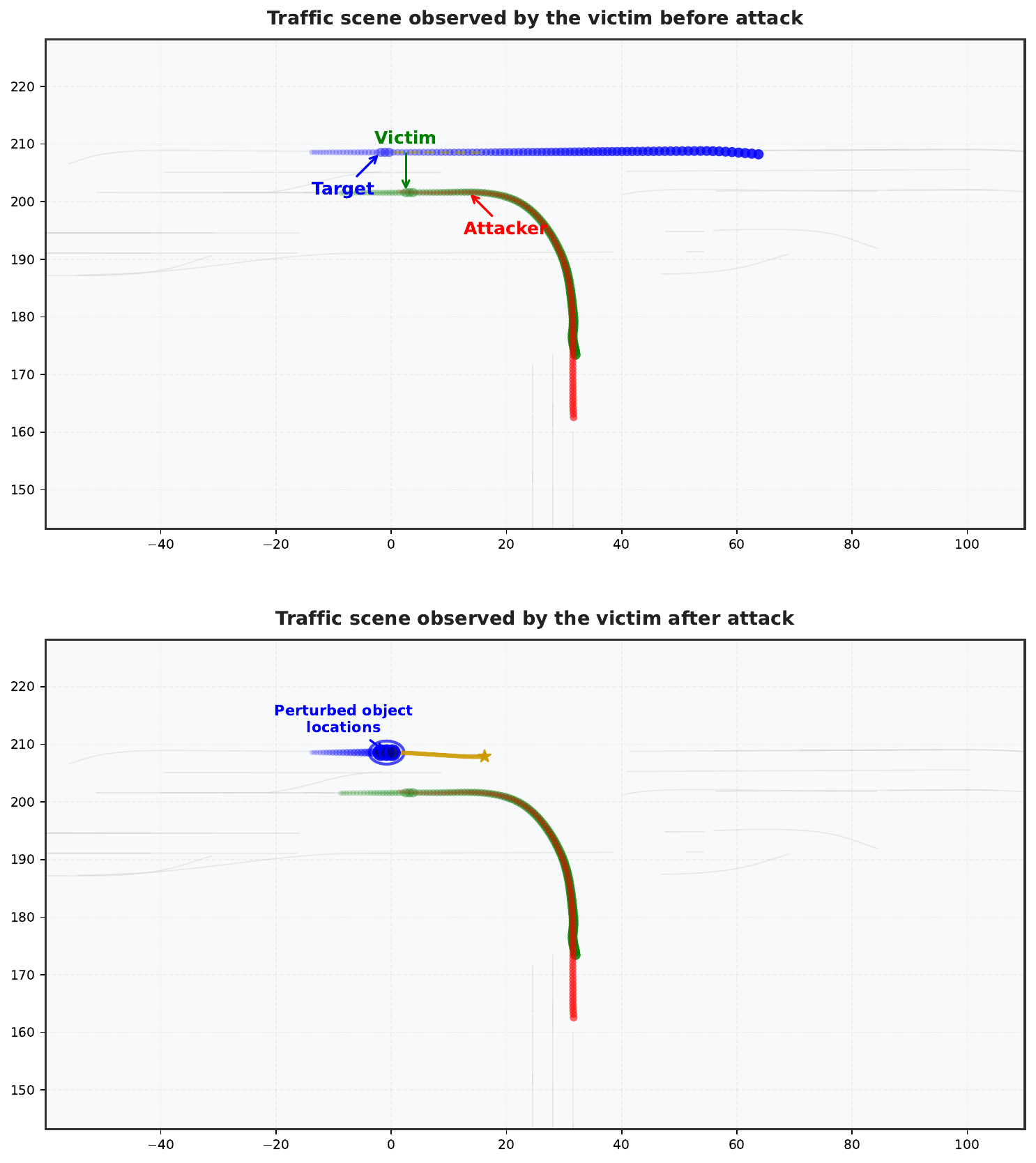}
\caption{Tracker absorption (fail)}
\end{subfigure}

\caption{Scenario attacks on AttFusion/OPV2V, victim BEV before (top of each pair) and after (bottom) the attack. Top row: successes. Bottom row: failure modes---(d)~target velocity orthogonal to the victim, (e)~prediction deflected away despite high ADE, (f)~tracker smoothing absorbs the perturbation.}
\label{fig:appendix_case_success}
\end{figure}

\section{Additional Case Studies}
\label{sec:appendix_case_studies}
\label{sec:appendix_carla_cases}

Figure~\ref{fig:appendix_case_success} extends \S\ref{sec:eval_case_study} on
AttFusion/OPV2V: green is the victim, blue the target, red the attacker. The successes span three
geometries: a same-lane trailing target shifted until the predictor infers a lane change
(minDist 2.26\,m, ADE 2.47\,m), a curved road carrying the prediction across the lane
boundary (3.17\,m, 1.78\,m), and an attacker turning in from a side road (2.95\,m, 0.96\,m).
Each failure has a distinct cause: a target crossing at $\sim$90$^\circ$ moves nearly
orthogonally, so the predictor extrapolates away regardless of the shift (7.65\,m minDist);
a distant target on a curve yields a large error curving \emph{away} from the victim
(6.64\,m ADE, 8.08\,m), so ADE alone does not capture success; and the Kalman tracker
absorbs a sub-voxel perturbation almost entirely (6.85\,m).

\begin{figure}[t]
\centering
\begin{subfigure}{0.15\textwidth}
\includegraphics[width=\textwidth]{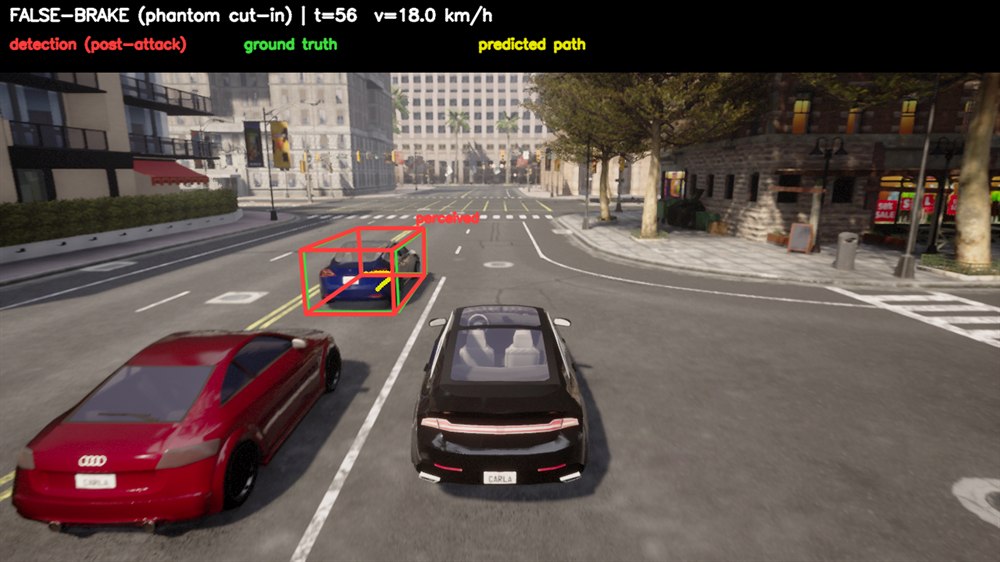}
\caption{Phantom cut-in $\rightarrow$ brake to 18.0\,km/h}
\end{subfigure}
\hfill
\begin{subfigure}{0.15\textwidth}
\includegraphics[width=\textwidth]{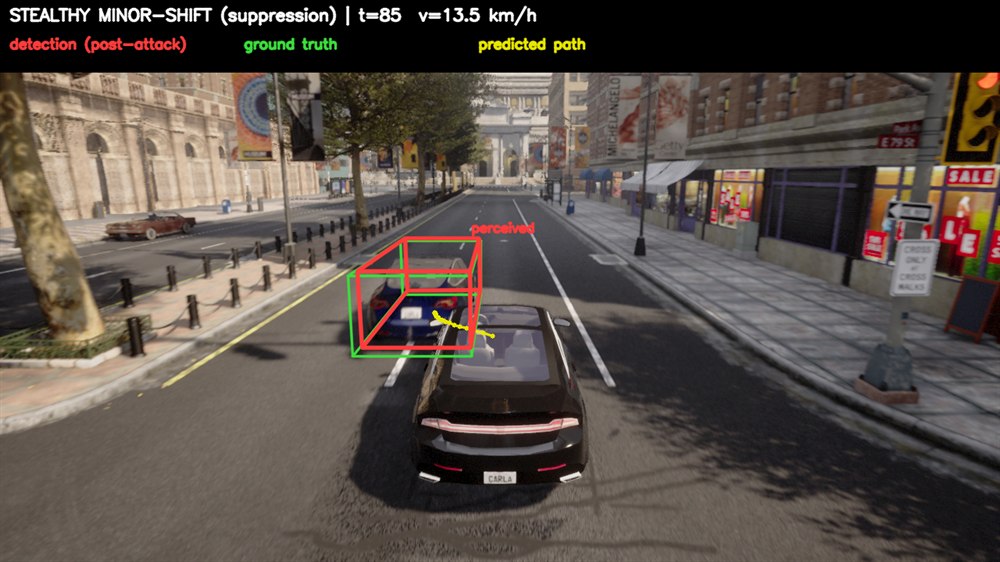}
\caption{Suppression $\rightarrow$ collision at 13.5\,km/h}
\end{subfigure}
\hfill
\begin{subfigure}{0.15\textwidth}
\includegraphics[width=\textwidth]{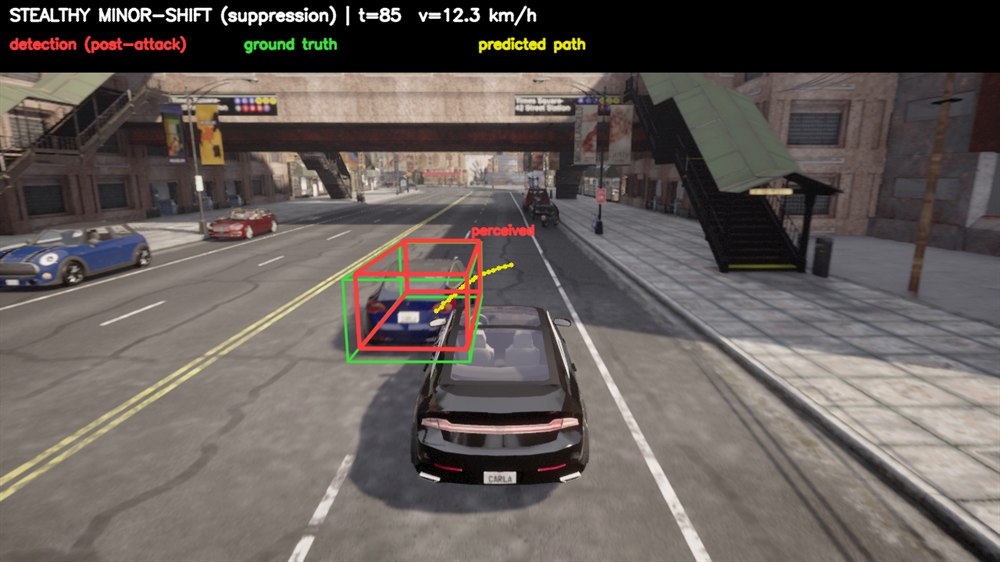}
\caption{Suppression $\rightarrow$ collision at 12.3\,km/h}
\end{subfigure}
\\[2pt]
\begin{subfigure}{0.15\textwidth}
\includegraphics[width=\textwidth]{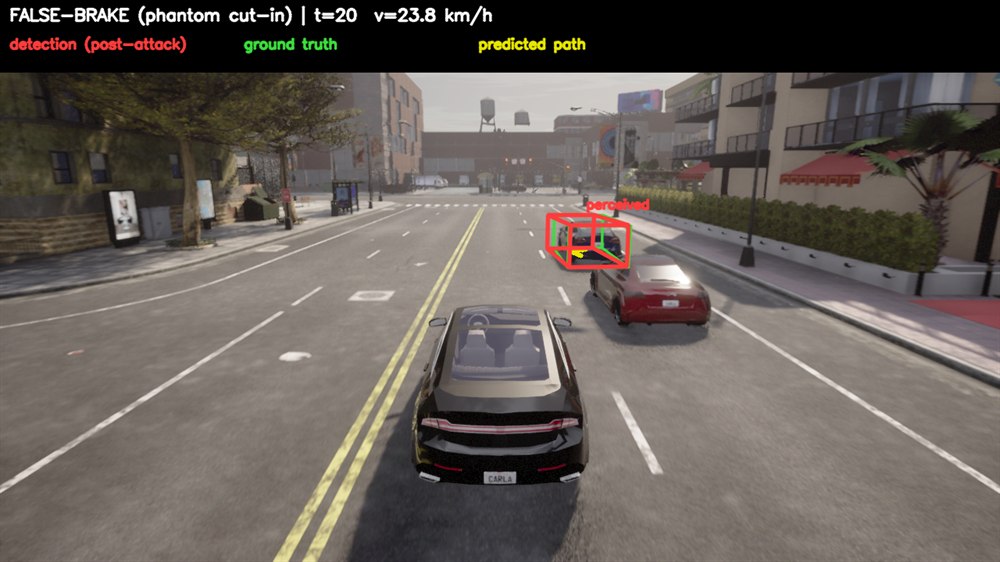}
\caption{Cut-in not gated as a lead (no braking)}
\end{subfigure}
\hfill
\begin{subfigure}{0.15\textwidth}
\includegraphics[width=\textwidth]{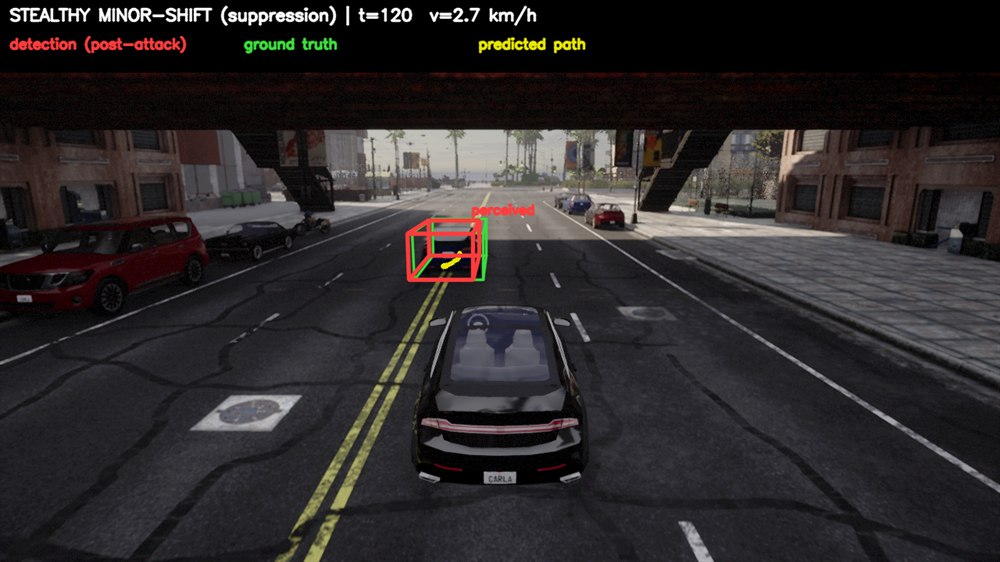}
\caption{Suppression absorbed $\rightarrow$ safe stop}
\end{subfigure}
\hfill
\begin{subfigure}{0.15\textwidth}
\includegraphics[width=\textwidth]{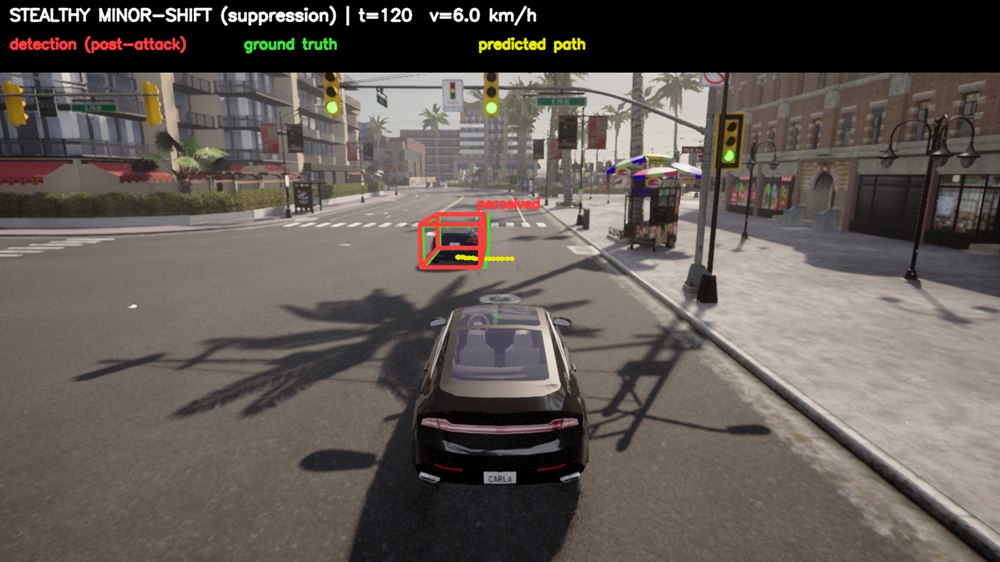}
\caption{Suppression with safety margin}
\end{subfigure}

\caption{Closed-loop CARLA cases. Top row: successes---(a)~a phantom cut-in induces unwarranted braking, (b,~c)~suppression of a marginal in-lane blocker causes collisions. Bottom row: failures---(d)~the cut-in is not gated as a close in-lane lead, (e)~the suppression shift is absorbed between attacked frames, (f)~the ego brakes with a comfortable margin.}
\label{fig:appendix_carla_success}
\end{figure}

Figure~\ref{fig:appendix_carla_success} extends \S\ref{sec:eval_closed_loop}, each panel a
chase-camera view with \textbf{red} the post-attack fused detection the ego acts on,
\textbf{green} the target's ground truth and \textbf{yellow} the GRIP++ predicted path.
A phantom cut-in brakes the ego from a $\ge 22$\,km/h cruise to 18.0\,km/h on a clear road,
and suppressing a marginal in-lane blocker produces impacts at 13.5 and 12.3\,km/h, in both
cases with the realized shift inside the sub-metre stealth bound. The failures show why the
attack raises the \emph{rate} of unsafe outcomes rather than guaranteeing one, and they stop
it at different points: a cut-in staying far enough ahead is never gated as a close in-lane
lead (ego holds 23.2\,km/h), so the planner's margins stop an attack that survived
perception and prediction; and since only alternate frames are perturbed, a blocker can
re-register in-lane on an unattacked one (minimum TTC 1.86\,s). Roads the attacker's
best-of-$N$ search flags as responders also do not always brake on a rerun.

\end{document}